\documentclass[12pt]{article}
\usepackage{amsmath,mathrsfs,amsthm,amsfonts}
\usepackage{graphicx}
\usepackage{enumerate}
\usepackage{natbib}
\usepackage{url} 
\usepackage{xcolor}
\usepackage{cleveref}
\usepackage{bm}
\usepackage{tikz}
\usetikzlibrary{trees}
\usetikzlibrary{shapes,decorations,arrows,calc,arrows.meta,fit,positioning}
\usetikzlibrary{shapes.multipart}
\tikzset{
	-Latex,auto,node distance =1 cm and 1 cm,semithick,
	state/.style ={ellipse, draw, minimum width = 0.7 cm},
	state1/.style ={ draw, minimum width = 0.7 cm},
	point/.style = {circle, draw, inner sep=0.04cm,fill,node contents={}},
	bidirected/.style={Latex-Latex,dashed},
	el/.style = {inner sep=2pt, align=left, sloped}
}

\newcommand{\blind}{1}

\def\pr{{\rm Pr}}
\def\E{{\rm E}}
\def\bmz{\bm{z}}
\def\bmZ{\bm{Z}}

\definecolor{dblue}{HTML}{0072B2}
\definecolor{dorange}{HTML}{D55E00}
\definecolor{dgreen}{rgb}{0.,0.6,0.}

\crefname{equation}{}{}

\crefname{theorem}{Theorem}{Theorems}

\crefname{lemma}{Lemma}{Lemmas}

\crefname{requirement}{Consideration}{Considerations}

\newtheorem{proposition}{Proposition}

\begin{document}

\def\spacingset#1{\renewcommand{\baselinestretch}%
  {#1}\small\normalsize} \spacingset{1}


\if1\blind
{
 \begin{center}
 	\spacingset{1.5} 
	{\Large\bf  Combining Broad and Narrow Case Definitions in  Matched Case-Control Studies: Firearms in the Home and Suicide Risk} \\ \bigskip \bigskip
	{\large  Ting Ye\footnote{Department of Biostatistics, University of Washington, \texttt{tingye1@uw.edu}.}, Kan Chen$^2$,  and Dylan S. Small\footnote{Department of Statistics and Data Science, The Wharton School, 
			University of Pennsylvania,  \texttt{dsmall@wharton.upenn.edu}.}
}
  \end{center}
} \fi

\if0\blind
{
  \bigskip
  \bigskip
  \bigskip
  \begin{center}
  	\spacingset{1.5} 
  {\Large\bf  Combining Broad and Narrow Case Definitions in  Matched Case-Control Studies: Firearms in the Home and Suicide Risk}
  \end{center}
  \medskip
} \fi

\bigskip
\begin{abstract}
{
	Does having firearms in the home increase suicide risk? To test this hypothesis, 
	a matched case-control study can be performed,  in which 
suicide case subjects are compared to living controls who are similar in observed covariates in terms of their retrospective exposure to firearms at home.  In this application, cases can be defined using a broad case definition (suicide) or a narrow case definition (suicide occurred at home). The broad case definition offers a larger number of cases but the narrow case definition may offer a larger effect size. Moreover, restricting to the narrow case definition may introduce selection bias (i.e., bias due to  selecting samples based on characteristics affected by the treatment) because exposure to firearms in the home may affect the location of suicide and thus the type of a case a subject is. We propose a new sensitivity analysis framework for combining broad and narrow case definitions in matched case-control studies, that considers the unmeasured confounding bias and selection bias simultaneously. We develop a valid randomization-based testing procedure using only the narrow case matched sets when the effect of the unmeasured confounder on receiving treatment and the effect of the treatment on case definition among the always-cases are controlled by sensitivity parameters. We then use the Bonferroni method to combine the testing procedures using the broad and narrow case definitions. {With the proposed methods, we find robust evidence that having firearms at home increases suicide risk.}
}

\end{abstract}

\noindent%
{\it Keywords:} {Causal inference}, {matching}, {observational study}, {selection bias}, {sensitivity analysis} 
\vfill

\newpage
\spacingset{1.5} 

\section{Introduction}

\label{sec: intro}
\subsection{Defining a Case in a Case-Referent Study}

\label{ssIntroConcept}

A case-control study, also known as a \emph{case-referent}  study, compares cases who have some disease (or rare event) to controls  (referents)   who do not have the disease, looking backward in time to contrast the frequency of treatment among cases and referents  \citep{Mantel:1973aa,  breslow1980, holland1987causal}.   In this article, we adopt the term ``referent'' rather than ``control'' because controls conventionally refer to the subjects who are untreated.

In many case-referent studies, there is some discretion in defining who is a case.  Aspects of defining a case include (i) how homogeneous should the cases be in terms of how similar their disease is \citep{Cole:1979aa}? (ii) when the disease ranges in severity, should cases be restricted to those with severe disease or include both severe and milder disease \citep{Lasky:1994aa, Fahmi:2008aa}? (iii) when there is the possibility of misclassifying a case, how much extra verification should be required to be a case \citep{Schlesselman::1982aa, BRENNER:1990aa,koepsell2003}? (iv) when there is information to suggest that a case's disease might have arisen from a source different than the treatment of interest, should this case be excluded \citep{JICK:1978aa}?   For each of these aspects of defining a case, we can characterize the choice as being between a {\it{broad}} vs. {\it{narrow}} case definition.  Whether a broad or narrow case definition is preferable depends on the setting.  A broad case definition offers a larger sample size of cases while a narrow case definition often offers a larger effect size because the effect is not diluted by having misclassified cases or cases who are not affected by the exposure.  For example, \citet{Cole:1979aa} and \citet{Acheson:1979aa} argue that for studying cancers of the body of the uterus, it is important to have a narrow case definition that excludes cancers of the cervix of the uterus, because exposures related to sexual activity have opposite effects on cancers of the body of the uterus compared to cancers of the cervix of the uterus and so lumping together these two histological types of cancers into one case definition would cancel out the effects of exposures.  In contrast,  \citet{Acheson:1979aa}  argues that for studying nasal cancers in boot/shoe workers, it is better to use a broad case definition that includes all histological types compared to a narrow case definition that focuses on only one histological type, because exposures tend to affect the whole spectrum of histological types among these workers and if only one histological type is considered, an effect of an exposure might be missed due to the small sample size.  Our goal in this paper is to provide statistical tools to aid in choosing between or combining  broad and narrow case definitions.

In choosing case definitions, we would like to (a) make the study valid in the sense that when there is no effect of treatment, the Type I error rate is what we promise and (b) make the study insensitive to bias in the sense that when the treatment does have an effect, the study has high power.  The type of power we seek in a case-referent study is {\it{power of sensitivity analysis}} which is motivated and defined as follows.  Because a case-referent study is an observational study, we are typically concerned about the possibility of unmeasured confounding.  We only find evidence for a treatment effect compelling if that evidence would remain even after conducting a sensitivity analysis that allows for some limited amount of unmeasured confounding.  Power of sensitivity analysis is the probability of there being evidence for a treatment effect in a sensitivity analysis that allows for a specified amount of unmeasured confounding;  see \citet[Chapter 14]{rosenbaum2004design, rosenbaum2010design}.  With the goals of (a) making sure the study is valid and (b) making the study as insensitive to bias as possible by maximizing the power of sensitivity analysis, we describe and illustrate tools for choosing case definitions and conducting sensitivity analyses with those case definitions.  The value of and methods for maximizing the power of sensitivity analysis has been discussed for cohort studies in epidemiology \citep{Zubizarreta:2013aa, Stuart:2013aa}.  In this paper, we develop such methods for case-referent studies.


\subsection{Firearms in the Home and Suicide Risk}
\label{subsec: motivate}

Suicide is a serious public health problem that can have lasting harmful effects on individuals, families, and communities \citep{CDC:suicide}. 
In the US in 2019, 50.4\% = 23,941/47,511 of suicides are committed using firearms \citep{cdc:2020}.  Approximately one in three U.S. households own guns \citep{schell2020state}.  Does having firearms at home cause an increase in suicide risk?

\citet{Wiebe:2003aa} examined this question using a {case-referent} analysis based on national samples of subjects 18 years or older. The suicide case subjects were obtained from the 1993 National Mortality Followback Survey (NMFS) \citep{nmfs1993}. The living referents were drawn from the 1994 National Health Interview Survey (NHIS) \citep{nhis1994}. The exposure information for the suicide cases was gathered by asking the descendant's next of kin or another person familiar with the decedent's life history the following question: ``At any time during the last year of life, were there any firearms kept in or around [the decedent's] home? Include those kept in a garage, outdoor storage area, truck or car.'' The exposure information for the living referents was based on a similar question: ``Are any firearms now kept in or around your home? Include those kept in a garage, outdoor storage area, truck or car.'' In both questions, the firearms include pistols, shotguns, rifles, and other types of guns, excluding guns that cannot fire, starter pistols, or BB guns.

{Of the 1,959 suicide cases ($ \geq 18 $ years old) in the NMFS, there are 1,365 suicide cases after excluding 30.3\%  with no information on firearms in the home.  These 1,365 suicide cases are our \emph{broad} cases.  After processing the NHIS data in the same way and excluding the subjects 95 years or older (as all the broad cases are 94 years or younger), there remains 97.2\% (19,190 out of 19,738) referents.} Each of the 1,365 broad cases is matched to one similar referent, controlling for sex, race, age, living arrangement (alone or not alone),  marital status, education, annual family income, military veteran status, geographical region, and population of the locality where the subject lived. The matching first stratifies on sex and race to ensure exact match on these two variables and then minimizes a robust Mahalanobis distance; see \citet[Chapter 8]{rosenbaum2010design} for discussion of the multivariate matching. Table \ref{tb1} describes covariate balance in the 1,365 matched pairs. {The matched samples show a good balance across covariates, with the exception of the missing living arrangement. However, as living arrangement is missing for only 2.4\% cases, this slight imbalance has a negligible impact.} {We also apply the sample-splitting classification permutation test by \cite{chen2023testing} to examine the quality of matching, and the results indicate that the matching is satisfactory.} Further detail about the datasets, and the construction of the matched sample are in the supplementary materials. {In the supplementary materials (Section C.2), we vary the details of matching and find very similar results.} Comparing with and without firearms at home, the odds ratio for suicide is 3.67. In a randomization test of no effect, the one-sided p-value is $1.9\times 10^{-62} $. This shows strong evidence that having firearms at home increases suicide risk, if controlling for the observed covariates by matching  has successfully rendered the matched subjects exchangeable. {However, as discussed in \cite{Wiebe:2003aa}, ``the greatest source of potential bias might be confounding from risk factors that were not measured or were controlled only partially.'' Such risk factors include 
mental health, and drug and alcohol use.} Sensitivity analysis is helpful for addressing this concern. A sensitivity analysis (described in Section \ref{sec: notation and review})
shows that the observed effect is insensitive to a bias of $ \Gamma= 3.9 $, where $ \Gamma= 3.9 $ means that two matched subjects may differ in their odds of having firearms at home by at most a factor of 3.9.  

To strengthen the causal conclusion to be more robust to possible unmeasured confounding, one may want to restrict to the suicides occurring at home -- the \emph{narrow} cases -- because ``If readily available firearms increase the risk of suicide, this effect should be most noticeable in an environment where guns are commonly kept''   \citep{Kellermann:1992aa}, and a larger effect tends to be harder to be explained away by unmeasured confounding bias. Looking at our 1,365 suicide cases, 1010 (74.0\%) took place in the home (our narrow cases), while the remaining consists of 293 (21.5\%) outside the home, and 65 (4.5\%) unknown. 
However, this seemingly reasonable approach may unintentionally introduce selection bias, because ``location (home) and exposure (gun in the home) are related, restricting the sample by location might have created bias...''   \citep{Wiebe:2003aa}.  More specifically, having firearms at home may induce a suicide case that would have occurred outside the home to commit suicide at home, so this subject would be included in our narrow case analysis if having firearms at home but be excluded if without firearms at home. Therefore, we are concerned with two types of bias in this analysis, the \emph{unmeasured confounding bias} that is due to the unobserved differences between the treated and untreated groups, and the \emph{selection bias} (also known as the collider bias \citep{hernan2020}) that is due to selecting samples based on characteristics affected by the treatment.

\vspace{-2mm}

\subsection{Prior Work and Our Contributions}

\citet{Rosenbaum:1991aa} develops sensitivity analysis methods for case-referent studies. When there is no gold standard  case definition, \citet{Small:2013aa} investigates the effect of different case definitions on sensitivity analysis when testing the null hypothesis of no treatment effect on both case definitions, while we are able to relax the null hypothesis and focus on testing the treatment effect for the broad case definition only.  
 \citet{Rosenbaum:2017aa} considers focusing on covariates-defining subgroups to improve the sensitivity analysis. However, both \citet{Small:2013aa} and \citet{Rosenbaum:2017aa} only consider selecting subjects based on characteristics unaffected by the treatment, which is intrinsically different from selecting subjects based on   characteristics affected by the treatment, as is the case discussed in Section \ref{subsec: motivate}. These two types of scenarios are also distinguished by \citet[Section 7.1.3]{rosenbaum2002} as the former does not introduce selection bias whereas the latter does. The latter type  is arguably more challenging and less investigated. The danger of selection bias has been noticed in recent applications; see, for example, \citet{KNOX:2020aa} and \citet{Schooling2021}. Statistical research on selection bias can be found at \citet{Rosenbaum:1984aa, Smith2019, Smith2020} and \citet{zhao2021note}. 

In this article, we address a specific cause of \emph{selection bias} -- the selection bias that is due to focusing on the narrow case definition -- and the unmeasured confounding bias simultaneously.  These biases, when  testing Fisher's sharp null hypothesis of no treatment effect based on the broad case definition,  can lead to an inflation of type I errors.  We propose a new sensitivity analysis framework for combining  the broad and narrow case definitions to strengthen the  causal conclusion against unmeasured confounding bias while accounting for the introduced selection bias. For testing Fisher's sharp null hypothesis of no treatment effect based on the broad case definition,  we develop a valid randomization-based testing procedure when restricting to the narrow case matched sets.  Intuitively, when the narrow case definition offers a larger effect size, restricting to the narrow case matched sets may increase the power of sensitivity analysis and thus strengthen the causal conclusion against unmeasured confounding. However, whether leveraging the narrow case definition can indeed be beneficial depends on the effect of the treatment on case definition among the always-cases, which is controlled by a sensitivity parameter $ \Theta $.  If a small $ \Theta $ suffices, then employing the narrow case definition can increase the power; otherwise, the adjustment of selection bias will counteract the larger effect size from the narrow case definition. This intuition is formalized from  analyzing the power of sensitivity analysis and design sensitivity. We combine the broad and narrow case definitions using the Bonferroni method, which attains the larger design sensitivity of the two case definitions and, unlike when used for combining multiple sensitivity analyses in other settings \citep{Rosenbaum:2009aa, Fogarty:2016aa}, is shown to be not unduly conservative when used for combining broad and narrow case definitions in our setting.

The rest of the article is organized as follows. Section \ref{sec: notation and review} introduces notation and setup, and reviews randomization inference and sensitivity analysis.  Section \ref{sec: main} proposes our new sensitivity analysis framework, a valid randomization-based test using the narrow case definition, and a combined test using both the broad and narrow case definitions. We return to the application in 
Section \ref{sec: application}. { \textsf{R}  code for the proposed method can be found in the  \textsf{R} package \textsf{broad.narrow}, which is posed at \textsf{http://github.com/tye27/broad.narrow.} \textsf{R} codes to replicate all the analyses are in the supplementary materials.}

\vspace{-4mm}

\section{Notation, Setup and Review}
\label{sec: notation and review}

\vspace{-2mm}
\subsection{Notations and Matched Case-Referent Studies}
\label{subsec: notation}
In the population of interest, there are $ I $ broad cases.  Each  broad case is matched with $ J-1 $ noncases or \emph{referents} with the same value of observed covariates. As such, in this case-referent study, we have formed $ I $ independent matched sets, with each set consisting of $J$ subjects and $J\geq 2$. Within matched set $ i =1,\dots, I$, the $ J $ subjects are numbered $ j=1,\dots, J $, where the first subject  is the $ i $th broad case without loss of generality.

For the $ j $th subject in matched set $ i $, we denote $ Z_{ij}= 1 $ if exposed to treatment and $ Z_{ij}=0 $ if not exposed to treatment.  Let $ m_i $ denote the number of exposed subjects in matched set $ i $, with $ 0\leq m_i\leq J $. {Let $r_{Tij}$ be the potential response if exposed to treatment and $r_{Cij}$ be the  potential response if not exposed to treatment, then the observed response is $R_{ij} = Z_{ij} r_{Tij} + (1-Z_{ij} ) r_{Cij}$.  For example, in Section \ref{subsec: motivate}, $r_{T{ij}}$ is a vector of the potential suicide outcome and location of suicide for individual $ij$ if having firearms at home, $r_{C{ij}}$ is a vector of the potential suicide outcome and location of suicide for individual $ij$ if having no firearms at home, and $R_{ij}$ is the observed vector of the suicide outcome and location of suicide for individual $ij$. 
The broad and narrow case definitions $\kappa_b(\cdot)$ and $\kappa_n(\cdot)$ are functions that map from a response to $\{0,1\}$, and  a narrow case must first be a broad case, i.e., $\kappa_n(\cdot)= 1  $ implies $\kappa_b(\cdot)= 1  $. Thus, $ \kappa_b(\cdot)= 0 $  denotes a referent, $ \kappa_b(\cdot)= 1 $  a broad case, $  \kappa_n(\cdot)= 1 $  a  narrow case, and $  \kappa_b(\cdot)=1 $ and $ \kappa_n(\cdot)=0 $   a  marginal case. With this notation, individual $ij$ has potential case variables $(\kappa_b(r_{Tij} ), \kappa_n(r_{Tij} ) ) $ if exposed to treatment, and potential case variables  $(\kappa_b(r_{Cij} ), \kappa_n(r_{Cij} ))$ if not exposed to treatment, so the observed case variables are  $(\kappa_b(R_{ij} ), \kappa_n(R_{ij} ))$.}  From our matching scheme, 
$ \kappa_b(R_{i1}) = 1 $ and $ \kappa_b(R_{i2}) = \dots= \kappa_b(R_{iJ}) = 0 $, for $ i=1,\dots, I $.

Each subject has an observed covariate vector $ x_{ij} $ used for matching  and an unobserved covariate $ u_{ij} $. Matching has  controlled the observed covariate in the sense that $ x_{ij}= x_{ij'} $ for $ 1\leq j<j'\leq J $ but matched subjects could differ in their unobserved covariate $ u_{ij} $. In our application, $x_{ij}$ consists of the covariates listed in Table \ref{tb1}. Although there are multiple important unmeasured confounders such as mental health, and drug and alcohol use, it is enough to consider a one-dimensional $u_{ij}$ with $0 \leq u_{ij}\leq 1$, namely the principal unobserved covariate \citep[Section 6.3]{rosenbaum2020modern}.

Let $ \mathcal{F} = \{(\kappa_b(r_{Tij}), \kappa_b(r_{Cij}), x_{ij}, u_{ij}  ), i=1,\dots, I, j=1,\dots, J\}$.  Write $ \mathcal{N} $ as the collection of matched set indexes where the broad case is also a narrow case, i.e.,  $ i\in \mathcal{N}  $ if and only if $ \kappa_n(R_{i1})=  \kappa_b (R_{i1})=1 $. 
Write $ \bmZ = (Z_{ij}, i=1,\dots, I, j=1,\dots, J)^T $ for the vector of dimension $ IJ $  containing the $ Z_{ij} $.    Let  $ \mathcal{Z}$ be the set of possible values  of $ \bmZ $ with $ \sum_{j=1}^J Z_{ij}=m_i $ for $ i=1,\dots, I $, and $ |\mathcal{Z}| = \prod_{i=1}^I \binom{J}{m_i} $, where $ |\mathcal{A}| $ denote the cardinality of a finite set $ \mathcal{A} $.  Conditioning on the event $ \bmZ\in \mathcal{Z} $ is abbreviated as conditioning on $ \mathcal{Z} $.  

\subsection{Hypothesis of Interest}
\label{subsec: hypothesis}

Throughout this article, we will be testing the Fisher's sharp null hypothesis of no treatment effect based on the broad case definition 
\begin{align}
    H_0: \kappa_b(r_{Tij}) =   \kappa_b(r_{Cij})  \text{ for all }  i,j , \label{eq: null hypothesis}
\end{align}  
regardless of whether we use the broad case definition, the narrow case definition, or both in the testing procedure.  When using the narrow case definition, it is important to note that under this null hypothesis $H_0$, the treatment does not change whether a subject is a broad case or not, but may still change whether a subject is a narrow case or a marginal case. Our interest in testing the $H_0$ in \eqref{eq: null hypothesis}  has a clear rationale in our application -- the hypothesis of interest concerns the effect of having firearms at home on the risk of suicide rather than its effect of moving suicidal subjects to commit suicide at home.

\subsection{Randomization Inference}
\label{subsec: randomization infererence}
If subjects are exchangeable within each matched set, i.e., there is no unmeasured confounding, for example, if the treatment assignment is random or if matching on $ x_{ij} $ has addressed the confounding issue,  then $\pr(\bmZ= \bmz \mid \mathcal{F}, \mathcal{Z}) = 1/ |\mathcal{Z}|$ for each $ \bm z\in \mathcal{Z} $.  

If Fisher's $ H_0 $ in \eqref{eq: null hypothesis} were true, it is straightforward to construct tests based on the broad case definition. Under $ H_0 $, $ \kappa_b(R_{ij})  = \kappa_b(r_{Cij}) $, where $ \kappa_b(r_{Cij})  $ is fixed by conditioning on $ \mathcal{F} $. We write $ Y_i = \sum_{j=1}^J Z_{ij} \kappa_b(R_{ij}) $ so $ Y_i =1 $ if the broad case in matched set $ i $ was exposed to the treatment and $ Y_i=0 $ otherwise. In case-referent studies, 
the common test of $ H_0 $ is the Mantel-Haenszel test which uses $ Y_b= \sum_{i=1}^I Y_{i}$, the number of broad cases exposed to the treatment \citep{mantel1959statistical, breslow1980, niven2012matched}. 
Exact inference can be obtained using the sum of independent extended hypergeometric distributions, when conditioning on $ \mathcal{F} $ and  $ \mathcal{Z} $; asymptotic inference can be obtained using the normal approximation. See \citet[Chapter 2.4]{rosenbaum2002} for details. 

\vspace{-2mm}
\subsection{Sensitivity Analysis in Observational Studies} 
\label{subsec: review SA}

In observational studies, matching may render subjects within each matched set nearly but not perfectly exchangeable, thus  matched subjects may still differ in their probability of receiving the treatment, i.e., $ \pr(\bmZ= \bmz \mid \mathcal{F}, \mathcal{Z}) \neq 1/ |\mathcal{Z}|$. In our application,  subjects who look similar may differ in their unobserved aspects such as mental health, and drug and alcohol use, and subjects with poorer mental health and higher levels of drug and alcohol use may have a higher probability of having firearms at home. If randomization inference assuming exchangeability rejects $ H_0 $, then a sensitivity analysis determines the departure from exchangeability that would lead to fail to reject $ H_0 $.

Assume independent  treatment assignments.  Rosenbaum's sensitivity model says 
\begin{align}
	\pr (Z_{ij}= 1\mid \mathcal{F} ) = {\rm expit} (\alpha_z(x_{ij}) + \log (\Gamma) u_{ij}) ~ \text{with} ~  0\leq u_{ij}\leq 1,  \label{eq: SA1}
\end{align}
where $ \alpha_z(\cdot) $ is an unknown real-valued function and $ \Gamma\geq 1 $ is an unknown sensitivity parameter. From model \eqref{eq: SA1}, two subjects with the same observed covariates $ x_{ij} $ and $ x_{ij'} $ may differ in their odds of receiving the treatment  by at most a factor of $ \Gamma $ \citep[Chapter 4.2.2]{rosenbaum2002}. Note that $ \Gamma=1 $  in \eqref{eq: SA1}   means no unmeasured confounding.

In a case-referent study under Fisher's $ H_0 $ and model \eqref{eq: SA1}, \citet{Rosenbaum:1991aa} shows that  the $ Y_i $'s are conditionally independent given $ \mathcal{F} $ and $ \mathcal{Z}$, with 
\begin{align}
	{\bar p}_{bi}=\frac{m_i}{m_i+ (J-m_i)\Gamma }\leq \pr \left( Y_{i} =1 \mid \mathcal{F}, \mathcal{Z} \right)\leq \frac{m_i \Gamma}{ m_i \Gamma +  J-m_i }= \bar{\bar p}_{bi}. \label{eq: SA broad}
\end{align}
The bounds are sharp in the sense that the upper bound is attained at $ \bm{u}_i= (1,0 \dots, 0) $ and the lower bound is attained at $ \bm{u}_i= (0,1 \dots, 1) $, where $ \bm u_i= (u_{i1},\dots, u_{iJ}) $.  Therefore, $ Y_b= \sum_{i=1}^I Y_i $ is the sum of $ I $ conditionally independent binary variables, with the probabilities bounded by \eqref{eq: SA broad}. Let $ \bar{\bar Y}_b $ be the sum of $ I $ independent binary variables, where the $ i $th variable takes the value 1 with probability $ \bar{\bar p}_{bi} $, and $ \bar Y_b $ be similarly defined with $ \bar p_{bi} $ in place of $ \bar{\bar p}_{bi} $. Under $ H_0 $,  the distributions of $ \bar{\bar Y}_b $ and $ \bar Y_b $ provide sharp bounds for the unknown  distribution of $ Y_b $, that is, for every $ k $, 
$$
	\pr (\bar Y_b \geq k  ) \leq 		\pr ( Y_b \geq k  \mid \mathcal{F}, \mathcal{Z}  ) \leq 		\pr ( \bar{\bar{Y}}_b \geq k  ). 
 $$ 
This yields sharp bounds on the p-value in the presence of a unmeasured confounding bias controlled by $ \Gamma $.  When $ \Gamma=1 $, the p-value upper and lower bounds become equal, and they reduce to the conventional p-value in a randomization inference 
reviewed in Section \ref{subsec: randomization infererence}.  

\vspace{-3mm}

\section{A Sensitivity Analysis Framework for Combining Broad and Narrow Case Definitions}
\label{sec: main}
\subsection{A Sensitivity Model}
\label{subsec: sensitivity model}
As reviewed in Section \ref{sec: notation and review}, use of the broad case definition to test  Fisher's $ H_0 $ in \eqref{eq: null hypothesis} is well studied. However, as motivated in Section \ref{sec: intro}, one may want to use the narrow case definition by restricting to the observed narrow case matched sets $  \mathcal{N}$, the matched sets satisfying $ \kappa_b(R_{ij}) =  \kappa_n(R_{ij}) $ for all $ j $,   to increase the power of sensitivity analysis.  For this purpose,  we use the Mantel-Haenszel test statistic restricting to the narrow case matched sets 
$Y_n=\sum_{i\in \mathcal{N}} Y_{i}= \sum_{i\in \mathcal{N}} \sum_{j=1}^{J} Z_{ij} \kappa_b(R_{ij}) $, where for $ i\in \mathcal{N} $,  $ \kappa_b(R_{ij})= \kappa_n(R_{ij}) $  and $ Y_{i}=1 $ if the narrow case in matched set $ i  $ was exposed to the treatment and $ Y_{i}=0 $  if otherwise. 

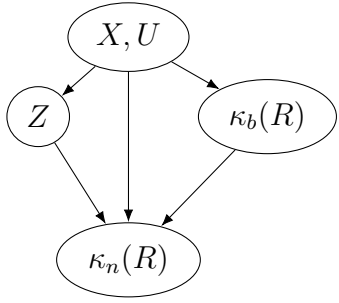
\begin{figure}[t]
	\centering
 \resizebox{!}{.25\textwidth}{
	\begin{tikzpicture}
		\node[state] (1) {$Z$};
		\node[state] (2) [below =of 1,xshift=1.2cm] {$\kappa_n(R)$};
		\node[state] (3) [right =of 1, xshift = 0.7cm] {$\kappa_b(R)$};
		\node[state] (5) [above =of 2, yshift=1cm] {$X, U$};
		\path (1) edge node[above] {} (2);
		\path (3) edge node[above] {} (2);
		\path (5) edge node[el,above] {} (2);
		\path (5) edge node[el,above] {} (3);
		\path (5) edge node[el,above] {} (1);
	\end{tikzpicture}} 
	\caption{A directed acyclic graph (DAG) under Fisher's $ H_0$ in \eqref{eq: null hypothesis}, where $ Z $ is the exposure, $ \kappa_b(R) $ and $ \kappa_n(R) $ are respectively the observed broad and narrow case status, $ X $ is the vector of measured covariates, $ U $ is the vector of unmeasured confounders. Restricting to the narrow case matched sets $ \mathcal{N} $  is equivalent with restricting to the matched sets satisfying  $ \kappa_b(R_{ij}) =  \kappa_n(R_{ij}) $ for all $ j $, and can create selection bias even under $ H_0 $.  \label{fig: dag}} 
\end{figure}

Restricting to the narrow case matched sets can introduce bias, even under Fisher's $ H_0 $ in \eqref{eq: null hypothesis},  because $  \kappa_n(R_{ij})  $ is a post-treatment variable that can be affected by the treatment (see the  directed acyclic graph in Figure \ref{fig: dag}). This is why the study by \citet{Kellermann:1993aa} is quite controversial; see, for example, the discussion in \citet[page 778]{Wiebe:2003aa}. 
Here, we clarify that if the goal were to test the null hypothesis of no effect on the narrow case definition (e.g., suicide at home), the conventional case-referent analysis using the
narrow case definition could be performed without introducing selection bias. However, as noted in Section 2.2, our goal is to test the effect of the treatment on the broad case definition (e.g., suicide) which makes selection bias a concern.

To develop a valid testing procedure for Fisher's $ H_0 $ in \eqref{eq: null hypothesis} using the narrow case definition, we first propose a new sensitivity analysis framework that considers the unmeasured confounding bias and selection bias simultaneously.   

Suppose that   $ Z_{ij} \perp (\kappa_b(r_{Tij}), \kappa_b(r_{Cij}) ,  \kappa_n(r_{Tij }), \kappa_n(r_{Cij })) \mid x_{ij}, u_{ij} $ for all $ i,j  $, and observations on distinct subjects are independent. We augment  Rosenbaum's sensitivity model \eqref{eq: SA1} by adding two  equations for the  narrow case definition:
\begin{equation} \label{eq: SA}
	\begin{split}
		\pi_{ij}:= 	&\pr(Z_{ij}=1\mid x_{ij}, u_{ij})= {\rm expit} (\alpha_z(x_{ij}) + \log (\Gamma) u_{ij}), \\
			\theta_{Tij}:=&  \pr(\kappa_n(r_{Tij})=1\mid  x_{ij}, u_{ij}, \kappa_b(r_{Cij}) =  \kappa_b(r_{Tij})= 1), \\
		\theta_{Cij}:= &\pr(\kappa_n(r_{Cij})=1\mid  x_{ij}, u_{ij}, \kappa_b(r_{Cij})=  \kappa_b(r_{Tij})= 1), \\
		\text{where }   &  0\leq u_{ij} \leq 1,   ~  1 \leq \theta_{Tij}/   \theta_{Cij} \leq \Theta, ~ \theta_{Cij}>0, \text{and } \Theta, \Gamma\geq 1.
	\end{split}
\end{equation}
We elaborate on the sensitivity model \eqref{eq: SA}. In model \eqref{eq: SA}, we have the sensitivity parameter $ \Gamma $ from Rosenbaum's  sensitivity model  \eqref{eq: SA1}.  When $ \Gamma=1 $, there is no unmeasured confounding.  We have an additional sensitivity parameter $ \Theta $ to bound $ \theta_{Tij}/   \theta_{Cij} $, the risk ratio of an \emph{always-case} being a narrow case with and without the treatment.  Here, the always-case is defined with respect to the broad case definition and refers to a subject who would be a broad case regardless of receiving the treatment or not. In our application, the always-cases are the subjects who would commit suicide regardless of having firearms at home or not. {As location (home) and exposure (gun in the home) are related \citep{Wiebe:2003aa}, in our application, it is likely that} 
 $ 1\leq \theta_{Tij}/\theta_{Cij} \leq \Theta$, which says that having firearms at home will make an always-case more likely and at most $ \Theta $ times more likely to commit suicide at home.   When $ \Theta=1 $, then under Fisher's $H_0$ in \eqref{eq: null hypothesis}, the treatment has no effect on the narrow case definition (i.e., $ \kappa_n(R_{ij}) \perp Z_{ij} \mid \mathcal{F}$ under $ H_0 $), thus restricting to the narrow case matched sets does not introduce any selection bias for testing $ H_0 $; otherwise, $ \Theta $ reflects the magnitude of selection bias.  
 Therefore, the sensitivity model \eqref{eq: SA} enables us to consider two sources of bias simultaneously: the unmeasured confounding bias controlled by $ \Gamma $ and the selection bias controlled by $ \Theta $. Notice that our results can be easily extended if instead we impose  $ \Theta^{-1} \leq \theta_{Tij}/\theta_{Cij}  \leq \Theta$ with $ \Theta\geq 1 $,  where  a  discussion is in the supplementary materials.

Moreover,  $u_{ij}$ is a unmeasured confounder that satisfies the unconfoundedness condition $ Z_{ij} \perp (\kappa_b(r_{Tij}), \kappa_b(r_{Cij}) ,  \kappa_n(r_{Tij }), \kappa_n(r_{Cij })) \mid x_{ij}, u_{ij} $. But this  $u_{ij}$  needs not modify  $\theta_{Tij}/\theta_{Cij}$. In fact, $\theta_{Tij}/\theta_{Cij}$ can be a constant no larger than $\Theta$.

Lastly,
$ \theta_{Tij}/\theta_{Cij} $ can be rewritten as 
\begin{align}
	\frac{\theta_{Tij}}{\theta_{Cij}}&= \frac{\pr(\kappa_n(R_{ij})=1\mid  x_{ij}, u_{ij}, \kappa_b(r_{Cij})=  \kappa_b(r_{Tij})= 1,  Z_{ij} = 1)}{\pr(\kappa_n(R_{ij})=1\mid  x_{ij}, u_{ij}, \kappa_b(r_{Cij})=  \kappa_b(r_{Tij})= 1, Z_{ij}=0 )}  \nonumber\\
	&  = \frac{\text{Odds} \{ \pr(Z_{ij}= 1\mid x_{ij}, u_{ij}, \kappa_b(r_{Cij})= \kappa_b(r_{Tij})= 1, \kappa_n(R_{ij}) = 1)\}  }{\text{Odds} \{ \pr(Z_{ij}= 1\mid x_{ij}, u_{ij}, \kappa_b(r_{Cij})=\kappa_b(r_{Tij})= 1)\}}\leq \Theta, \label{eq: SA2} 
\end{align}
where $ \text{Odds} \{p\} = p/(1-p)$. Thus, $ \theta_{Tij}/\theta_{Cij} $ is the odds ratio of  receiving treatment for the same subject with and without 
the information that this subject is a narrow case, which is different from the odds ratio of receiving treatment for two matched subjects in Rosenbaum's sensitivity model  \eqref{eq: SA1}.

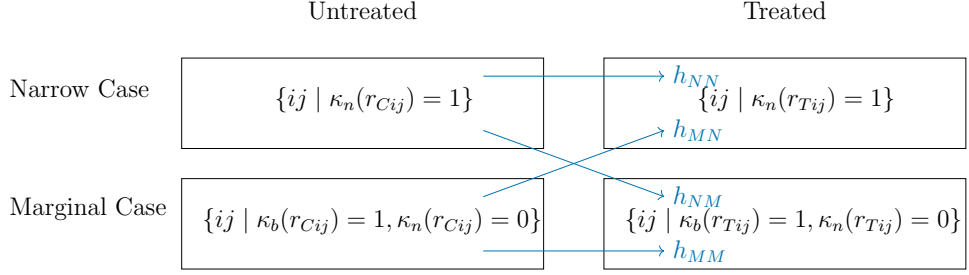
\begin{figure}[t]	
	\centering
	\resizebox{!}{!}{
		\begin{tikzpicture}
			\draw[draw=black] (10,5) rectangle ++(6,1.5);
			\draw[draw=black] (17,5) rectangle ++(6,1.5);
			\draw[draw=black] (10,3) rectangle ++(6,1.5);
			\draw[draw=black] (17,3) rectangle ++(6,1.5);
			\node[above] at (13,7) {Untreated};
			\node[above] at (20,7) {Treated};
			\node[right] at (7,6) {Narrow Case};
			\node[right] at (7,4) {Marginal Case};
			\node[right] at (10.2,3.8) {$ \{ij\mid \kappa_b(r_{Cij})= 1, \kappa_n(r_{Cij})= 0 \} $ };
			\node[right] at (17.2,3.8) {$ \{ij\mid \kappa_b(r_{Tij})= 1, \kappa_n(r_{Tij})= 0 \} $ };
			\node[right] at (11.4,5.8) {$ \{ij\mid \kappa_n(r_{Cij})= 1 \} $ };
			\node[right] at (18.4,5.8) {$ \{ij\mid \kappa_n(r_{Tij})= 1 \} $ };
			\draw[->,dblue] (15,4.2) -- (18,5.3) node[pos=1,right] {$ h_{MN} $};
			\draw[->,dblue] (15,5.3) -- (18,4.2) node[pos=1,right] {$ h_{NM} $};
			\draw[->,dblue] (15,6.2) -- (18,6.2) node[pos=1,right] {$ h_{NN} $};
			\draw[->,dblue] (15,3.3) -- (18,3.3) node[pos=1,right] {$ h_{MM} $};
	\end{tikzpicture}}
	\caption{An illustration of how the cases are moved by the treatment among the always-cases, within the stratum defined by levels of $( x_{ij}, u_{ij}) $ in which  
		$ \theta_{Tij}/\theta_{Cij} $ is the largest.  The left panel is the case allocation when all the subjects
		are untreated, and the right panel is  the case allocation when all are treated.  The numbers $ h_{NN}, h_{MN}, h_{NM}, h_{MM} $ are expected values. 
		\label{fig: 1}}
\end{figure}

The next question is how do we interpret $ \Theta $? Consider the stratum defined by levels of $( x_{ij}, u_{ij}) $ in which  
$ \theta_{Tij}/\theta_{Cij} $ is the largest. Figure \ref{fig: 1} illustrates how the cases are moved by the treatment among the always-cases. The left panel is the case allocation when all the subjects
are untreated, the right panel is  the case allocation when all the subjects are treated. Figure \ref{fig: 1} shows that $ h_{NN} $ narrow cases and $ h_{MM} $ marginal cases are not affected by the treatment, and the treatment  moves $ h_{MN} $ marginal cases to be narrow cases and moves $ h_{NM} $ narrow cases to be marginal cases, where $ h_{NN}, h_{MN}, h_{NM}, h_{MM} $ are expected values.  Then 	 $ \theta_{Tij}/\theta_{Cij} \leq\Theta $ requires  $   (h_{NN}+h_{MN}) /(h_{NN}+h_{NM})\leq \Theta$, which can be written as $ (h_{MN}-h_{NM}) / (h_{NN}+h_{MN} ) \leq 1-1/\Theta$.  When it is reasonable to assume that $ h_{NM} $ is small compared to $ h_{MN} $, this can be further approximated by
$ h_{MN} / (h_{NN}+h_{MN} ) \leq 1-1/\Theta$. {Hence, in the context of our application,  $ \Theta=1.22 $ means that among the always-cases,  up to $ 1-1/1.22= 18\%$ of the subjects committing suicide at home when having firearms at home would commit suicide in other places if  without firearms at home; $ \Theta=1.36 $ would allow up to $ 1-1/1.36= 26\%$ such subjects.
Still, $\Theta$ bounds $\theta_{T_ij}/\theta_{Cij}$ that are defined among always-cases, which are never observed. In our application in Section \ref{sec: application}, we provide an informal benchmark approach. Future research requires more formal calibration strategies to enhance the interpretability of $\Theta$; see \cite{Imbens2003}, \cite{hsu2013calibrating}, \cite{cinelli2020making}, and \cite{zhang2020calibrated} for some works on the calibration of sensitivity analysis.
}

\vspace{-5mm} 

\subsection{Testing}

To develop a valid randomization-based testing procedure for Fisher's $ H_0 $ using only the narrow case matched sets, we first study  $   \pr \left( Y_{i} =1 \mid i\in \mathcal{N}, \mathcal{F}, \mathcal{Z} \right)= \pr \left( Z_{i1} =1 \mid i\in \mathcal{N}, \mathcal{F}, \mathcal{Z} \right)$. This quantity differs from its broad case definition counterpart $  \pr \left( Y_{i} =1 \mid \mathcal{F}, \mathcal{Z} \right) =\pr \left( Z_{i1} =1 \mid \mathcal{F}, \mathcal{Z} \right) $ in \eqref{eq: SA broad} because conditioning on $ i\in \mathcal{N} $ changes the probability that the case receives the treatment. The following proposition places bounds on the unknown probability $   \pr \left( Y_{i} =1 \mid i\in \mathcal{N}, \mathcal{F}, \mathcal{Z} \right) $. 

\begin{proposition} \label{proposition: Yi}
	Under Fisher's null in \eqref{eq: null hypothesis} and the sensitivity model \eqref{eq: SA} with fixed $ \Theta$  and  $\Gamma $,  then
	\begin{align}
		{\bar p}_{ni} = \frac{m_i}{m_i+ (J-m_i)\Gamma }\leq \pr \left( Y_{i} =1 \mid i\in \mathcal{N}, \mathcal{F}, \mathcal{Z} \right)\leq \frac{m_i \Theta\Gamma}{ m_i \Theta\Gamma +  J-m_i} = \bar{\bar p}_{ni}. \label{eq: prob bound}
	\end{align}
	The bounds are sharp in the sense that  the lower bound is attained at $ \bm{u}_i= (0,1 \dots, 1) $ and $ \theta_{Ti1}/\theta_{Ci1} = 1 $, and the upper bound is attained at $ \bm{u}_i= (1,0 \dots, 0) $ and  $ \theta_{Ti1}/\theta_{Ci1} = \Theta $. 
\end{proposition} 
{The proof is in the supplementary materials, which extends the proof in \cite{Rosenbaum:1991aa} to our new sensitivity model in \eqref{eq: SA}.

Proposition \ref{proposition: Yi} clearly demonstrates that restricting to narrow case matched sets can introduce bias and inflate the type I error rate. Consider a simple scenario where $\Gamma=1$ (i.e., no unmeasured confounding bias). Under Fisher's $H_0$ and when considering all the broad case matched sets, from \eqref{eq: SA broad}, the broad case and the referents have an equal probability of being treated, which is the basis for the validity of the randomization-based test using the broad case definition. However, from \eqref{eq: prob bound}, we see that even under Fisher's $H_0$, the narrow case has a higher probability of being treated compared to the referents. Consequently, applying a randomization-based test while restricting to the narrow case definition will lead to over-rejection. In our application, this means that even under the null hypothesis that having firearms at home has no effect on suicide, because firearms can make suicides more likely to occur at home, having firearms will be associated with suicide at home. Therefore, a direct application of the randomization-based test while restricting to narrow case matched sets will result in inflated type I error rates.  
}



Now consider the narrow case test statistic $Y_n= \sum_{i\in \mathcal{N}}  Y_i  $, which is the sum of $ |\mathcal{N}| $ conditionally independent binary random variables taking  value 1 with probabilities bounded by $ {\bar p}_{ni}  $ and $ \bar{\bar p}_{ni}  $.  Let $ \bar{\bar Y}_n $ be the sum of independent binary random variables, each taking  value 1 with probability $ \bar{\bar p}_{ni} $, and define $ {\bar Y}_n $ in the same way but with  $ {\bar p}_{ni}$ in place of $ \bar{\bar p}_{ni} $.  The next proposition gives known bounds on the unknown distribution of $ Y_n $ and $ Y_b $, which is an immediate consequence of Proposition \ref{proposition: Yi} and  the results reviewed in Section \ref{subsec: review SA}.

\begin{proposition} \label{prop: SA bounds}
	Under the same conditions in Proposition \ref{proposition: Yi},  then for the Mantel-Haenszel statistics $ 	Y_n = \sum_{ i\in \mathcal{N}} Y_{i}  $ and $ 	Y_b =  \sum_{i=1}^{I} Y_{i}  $, 
	\begin{align*}
		&\pr (\bar Y_n \geq k  ) \leq 		\pr ( Y_n \geq k  \mid   \mathcal{N},  \mathcal{F}, \mathcal{Z}  ) \leq 		\pr ( \bar{\bar{Y}}_n \geq k  ), \\
		&\pr (\bar Y_b \geq k  ) \leq 		\pr ( Y_b \geq k  \mid \mathcal{F}, \mathcal{Z}  ) \leq 		\pr ( \bar{\bar{Y}}_b \geq k  ),
	\end{align*}
	where conditioning on the narrow case matched sets is abbreviated as conditioning on $ \mathcal{N} $.   The bounds are sharp in the sense that  both the upper bounds  can be attained at $ \bm{u}_i= (1,0 \dots, 0) $ and  $ \theta_{Ti1}/\theta_{Ci1} = \Theta $,   both the lower bound can be attained  at $ \bm{u}_i= (0,1 \dots, 1) $ and $ \theta_{Ti1}/\theta_{Ci1} = 1 $, for $ i=1,\dots, I $. 
\end{proposition}
In our application, we consider a one-sided test. Then, the p-value upper bounds when using either the broad case or  narrow case  are achieved when for every matched set, the case has $ u_{i1}  =1 $, and thus a higher probability of having firearms at home. In the meantime, having  firearms at home will make the case $ \Theta $ times more likely to commit suicide at home compared to having no firearm at home. This is the situation in which having firearms at home (the exposure) and committing suicide at home (the narrow case) exhibits the strongest association when in fact having firearms at home has no effect on committing suicide (i.e., under Fisher's $ H_0 $).

The bounds in Proposition \ref{prop: SA bounds} are exact, and their exact values can be easily obtained by convolving probability generating functions  \citep[Chapter 3.9]{rosenbaum2010design}. When $ |\mathcal{N}| $ is reasonably large, we can also calculate the tail probability bounds using the large-sample normal approximation 
\begin{align}
	\frac{{\bar Y}_n - \sum_{i\in \mathcal{N}} {\bar p}_{ni}}{\sqrt{\sum_{i\in \mathcal{N}}  {\bar p}_{ni} (1- {\bar p}_{ni}) }}\xrightarrow{d} N(0,1), \qquad \frac{\bar{\bar Y}_n - \sum_{i\in \mathcal{N}} \bar{\bar p}_{ni}}{\sqrt{\sum_{i\in \mathcal{N}}  \bar{\bar p}_{ni} (1- \bar{\bar p}_{ni}) }}\xrightarrow{d} N(0,1), 
	\label{eq: bound n}
\end{align}
and 
\begin{align*}
	1- \Phi \left( \frac{k- \sum_{i\in \mathcal{N}} {\bar p}_{ni} }{\sqrt{\sum_{i\in \mathcal{N}}  {\bar p}_{ni} (1- {\bar p}_{ni})}} \right) \leq 	\pr ( Y_n \geq k  \mid   \mathcal{N},  \mathcal{F}, \mathcal{Z}  )   \leq  1- \Phi \left( \frac{k- \sum_{i\in \mathcal{N}} \bar{\bar p}_{ni} }{\sqrt{\sum_{i\in \mathcal{N}}  \bar{\bar p}_{ni} (1- \bar{\bar p}_{ni})}} \right), 
\end{align*}
where $ \Phi(\cdot) $ is the standard normal cumulative distribution. The normal approximation for using the broad case definition can be obtained in the same way. 

Both the broad case test and the narrow case test are conditional inference procedures which only exploit the randomness of treatment assignments. Compared to the broad case test,  the narrow case test further conditions on the observed narrow case matched sets $ \mathcal{N} $. To motivate this additional conditioning step, we first note that there are no narrow cases in	marginal case matched sets $ \mathcal{N}^c $, and thus these matched sets do not contribute a stochastic component to the Mantel-Haenszel test based on the narrow case definition. Therefore, this conditional step is to focus on the ``discordant'' matched sets that contribute to the randomization inference.

With  two valid tests for Fisher's null  $ H_0 $, we can combine them using the Bonferroni method. Write  the one-sided p-values as $  \text{Pval}_n  =	\pr ( Y_n \geq Y_{n,\text{obs}}  \mid   \mathcal{N},  \mathcal{F}, \mathcal{Z}  )  $ and $ \text{Pval}_b  =	\pr ( Y_b \geq Y_{b,\text{obs}}   \mid \mathcal{F}, \mathcal{Z}  ) $, where $ Y_{n,\text{obs}}  $ and $ Y_{b,\text{obs}}  $ are the two test statistics.
We reject $ H_0 $ when the Bonferroni p-value is below the significance level $ \alpha $, i.e.,  \vspace{-2mm}
\begin{align*}
	\alpha \geq \underbrace{2 	\min \{ \max_{\bm u} \text{Pval}_b, \max_{\bm u, \bm \theta} \text{Pval}_n  \}}_{\text{ Bonferroni  p-value}},
\end{align*}
where $ \bm u= (\bm u_1, \dots,\bm u_I) $ and $ \bm\theta= (\theta_{Tij}, \theta_{Cij}, i =1,\dots, I, j=1, \dots, J) $.

In many applications, it would be difficult to know whether $Y_n$ or $Y_b$ has higher power before examining the data.  It is in these situations that using the Bonferroni method to combine the broad and narrow case definitions is especially attractive. This is because in the limit, as the sample size goes to infinity, the combined test has the same insensitivity to unmeasured confounding bias as the better one of the broad case test and the narrow case test; in finite samples, the
combined test pays a small price in terms of power compared to knowing a priori which
case definition is better, but gains a lot compared to making an incorrect guess about which
case definition is better. 
More details are in  Section B of the Supplementary Materials, where we compare the finite-sample power of sensitivity analysis and design sensitivity based on $Y_b$, $Y_n$, and the combined test \citep[Chapter 14]{rosenbaum2004design, rosenbaum2010design}. 

As a final comment for this section, the Bonferroni method has a reputation of being quite conservative when used to combine multiple sensitivity analyses \citep{Rosenbaum:2009aa, Fogarty:2016aa}, because  the  worst-case p-values for different tests may be attained at different allocations of the unmeasured confounders. More specifically,  if we fail to reject $ H_0 $ because the Bonferroni p-value exceeds $ \alpha $, it could be the case  that there does not exist a single allocation of the unmeasured confounders which simultaneously makes all the component tests insignificant, and thus we should have rejected $ H_0 $. 
However, this is not a concern in our setting because the worst-case p-values for the  broad and narrow case definitions are attained at the same $ \bm u $ and $ \bm \theta $ as noted in Proposition \ref{prop: SA bounds}.  Therefore,  the Bonferroni method is not unduly conservative in our setting.

\section{Application: Firearms at Home and Suicide Risk}
\label{sec: application}

\subsection{Results from the Matched Case-Referent Study}
\label{subsec: application, referent group 1}
Recall from Section \ref{subsec: motivate} that we formed 1,365 case-referent  matched pairs to investigate whether having firearms in the home increases suicide risk. Comparing with and without firearms at home, the odds ratio for suicide is 3.67.  In a randomization test of Fisher's null in \eqref{eq: null hypothesis}, that is, whether having firearms at home or not would not change the subjects' suicide case status, the one-sided p-value is $1.9 \times 10^{-62}$, showing strong evidence that having firearms at home increases  suicide risk. This effect is robust to an unmeasured confounding bias of $ \Gamma=3.9 $, meaning that two matched subjects differing in their odds of having firearms at home by at most a factor of 3.9 would not be able to explain away the observed effect. {To provide some informal benchmark, we might compare $\Gamma=3.9$ to the exponentiated coefficients of the observed covariates in the propensity score of having firearms at home \citep{yu2021information}. Among the binary covariates in Table \ref{tb1}, the largest exponentiated coefficients are: 5.26 for white race (compared to other race) and 2.70 for non-MSA (compared to population $\geq$ 250,000). Therefore, although $\Gamma=3.9$ is already a fairly large departure from exchangeability, it is not entirely implausible.}

As motivated in Section \ref{subsec: motivate},  leveraging the narrow case definition may potentially strengthen the causal conclusion to be more robust against unmeasured confounding.  Note that out of the 1,365 broad case matched sets, there are 1,010 narrow case matched sets -- the matched sets with the case committing suicide at home.  Indeed, the odds ratio of committing suicide at home with and without firearms at home is 4.50, larger than 3.67, the counterpart for the broad case definition.

\begin{figure}
	\centering
	\includegraphics[scale=0.9]{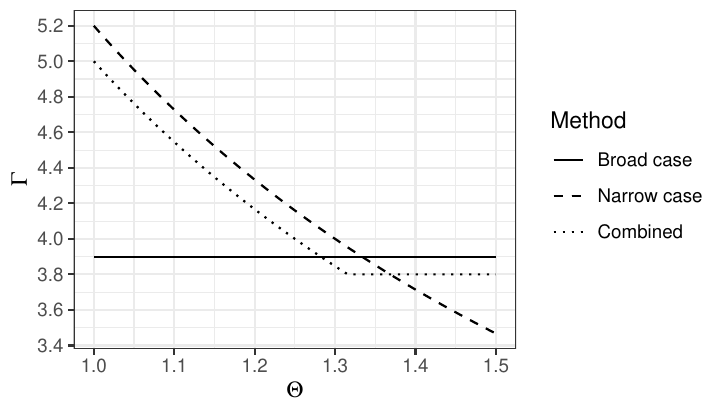}
	\caption{ The largest $ \Gamma $ against different values of the sensitivity parameter $ \Theta $, under which we are still able to conclude that there is a significant treatment effect of having firearms at home on suicide. \label{fig: data}}
\end{figure}

We apply the proposed sensitivity analysis from using the narrow case test and the combined test developed in Section \ref{sec: main}. The results are in 
Figure \ref{fig: data}, where we plot the largest $ \Gamma $ against different values of $ \Theta $, under which we are still able to conclude that there is a significant treatment effect of having firearms at home on suicide.
If there is no selection bias (i.e., $\Theta=1$), an unobserved covariate that increases the odds of having firearms at home by more than fivefold ($\Gamma= 5.2$) could not explain the observed association. If $\Theta=1.22$  suffices to control the effect of treatment on case definition among the always-cases, then the narrow case test leads to $\Gamma= 4.2$ and thus leveraging the narrow case definition still strengthens the causal conclusion against unmeasured confounding. If we keep increasing $\Theta$ to be equal to 1.36, then we have $\Gamma=3.8$ and so the narrow case test becomes more sensitive to unmeasured confounding compared to the broad case test.  

{Here, we provide some informal benchmarks to enhance the interpretability of $\Theta$.  
From the sensitivity model in   \eqref{eq: SA}, $\Theta$ is an upper bound of  
$$
\frac{\theta_{Tij}}{\theta_{Cij}}  = \frac{\pr(\kappa_n(r_{Tij})=1\mid x_{ij}, u_{ij}, \kappa_b(r_{Cij})=  \kappa_b(r_{Tij})= 1)}{\pr(\kappa_n(r_{Cij})=1\mid x_{ij}, u_{ij},  \kappa_b(r_{Cij})=  \kappa_b(r_{Tij})= 1 )}, 
$$
which is defined among the always-cases and can never be observed. Suppose that 
\emph{case monotonicity} holds  (i.e., $ \kappa_n(r_{Tij}) \geq  \kappa_n(r_{Cij})$ and $ \kappa_b(r_{Tij}) \geq  \kappa_b(r_{Cij})$ for every $ i,j $), meaning that if a subject would die from suicide (at home) without firearms at home, the subject would also die from suicide (at home) if they  had firearms at home. Suppose also that  
\begin{align}
\begin{split}
   	& \pr (\kappa_n (r_{Tij}) = 1\mid  x_{ij}, u_{ij}, \kappa_b (r_{Tij})= 1,  \kappa_b (r_{Cij})= 0 )\\
  & \geq 	\pr (\kappa_n (r_{Tij}) = 1\mid x_{ij}, u_{ij}, \kappa_b (r_{Tij})= 1,  \kappa_b (r_{Cij})= 1)\nonumber, 
\end{split}
\end{align}
where the left-hand side term  is the probability of suicide at home if having firearms at home among subjects whose suicides are attributable to having firearms at home. This left-hand side probability is likely to be very high because if a subject would only commit suicide if the subject had firearms at home, it is very likely that the subject would commit the suicide at home if firearms were at home.  
Therefore, it is reasonable to assume that the left-hand side term is no less then the probability on the right-hand side, which is the probability of suicide at home if having firearms at home  among always-cases. Under these two aforementioned conditions, we can bound $\theta_{Tij}/\theta_{Cij}$  as follows, 
\begin{align*}
	& \frac{\pr (\kappa_n (R_{ij}) = 1\mid  x_{ij}, u_{ij}, \kappa_b (R_{ij})= 1, Z_{ij} = 1)  }{ 	\pr (\kappa_n (R_{ij}) = 1\mid x_{ij}, u_{ij},  \kappa_b (R_{ij})= 1, Z_{ij}=0)  } =  \ \frac{\pr (\kappa_n (r_{Tij}) = 1\mid  x_{ij}, u_{ij}, \kappa_b (r_{Tij})= 1)  }{ 	\pr (\kappa_n (r_{Cij}) = 1\mid x_{ij}, u_{ij},  \kappa_b (r_{Cij})= 1)  }  
    \geq   \frac{\theta_{Tij}}{\theta_{Cij}}.
\end{align*}
Although the first term remains unidentifiable due to the presence of $u_{ij}$, it can potentially be identified in the future if $u_{ij}$ is measured. For now, we can calibrate it to observed covariates by considering 
\begin{align}
    \frac{\pr (\kappa_n (R_{ij}) = 1\mid  x_{ij}, \kappa_b (R_{ij})= 1, Z_{ij} = 1)  }{ 	\pr (\kappa_n (R_{ij}) = 1\mid x_{ij},  \kappa_b (R_{ij})= 1, Z_{ij}=0)} .\label{eq: ratio}
\end{align}
This can be done by using fitted values from a logistic regression model of $\kappa_n(R_{ij})$ on all observed covariates and treatment among the broad cases. From the boxplot of this estimated quantity in
Figure \ref{fig: ratio}, we see that the 90\% percentile is 1.36, which shows that $\theta_{Tij}/\theta_{Cij}$ may be bounded below 1.36. It is important to note that this benchmark approach has limitations. First,  the quantity in \eqref{eq: ratio} is an approximation to an upper bound because we can not adjust for $u_{ij}$. To assess the extent of this approximation due to not adjusting for a confounder, we have tried dropping one of the observed covariates and found that the 90\% percentiles remain close to 1.36. Second, the values shown in Figure \ref{fig: ratio} have estimation errors that were not accounted for. Formal approaches to calibrate the sensitivity parameter $\Theta$ are needed in future research. 

 \begin{figure}[t]
	\centering
	\includegraphics[scale=.7]{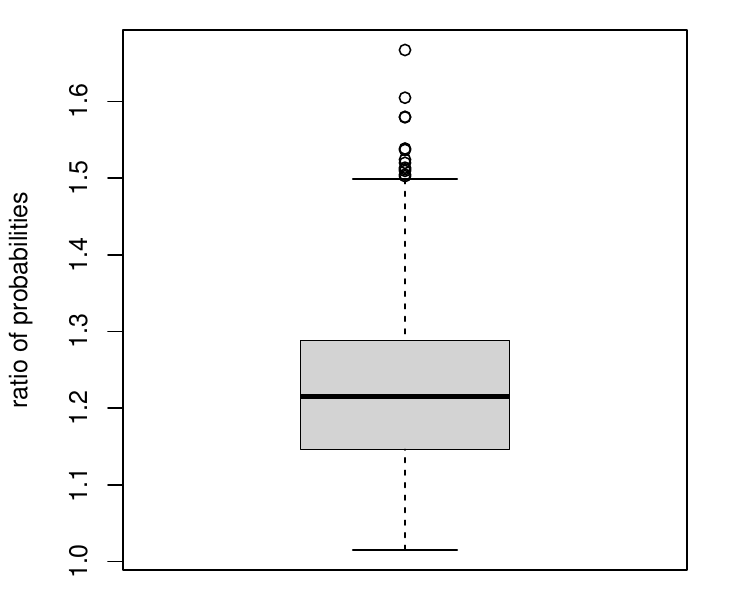}
	\caption{Boxplots of the estimated ratio defined in \eqref{eq: ratio}.\label{fig: ratio}}
\end{figure}

When we do not know a priori whether the broad case test or the narrow case test will be more robust to unmeasured confounding bias, the combined test usually performs well. From Figure \ref{fig: data}, we see that it has a small loss in $\Gamma$ compared to the more robust method, but results in a large gain in $\Gamma$ compared to the less robust method.



Lastly, as a comparison, we apply the standard logistic regression approach that adjusts for all observed covariates listed in Table \ref{tb1}. Comparing with and without firearms at home and conditional on all observed covariates, the odds ratio for suicide is 4.48, for suicide at home when restricting to the narrow case matched sets is 5.92. Note, however, that these are estimates of conditional odds ratios, which differ from unconditional odds ratios obtained  through unconditional logistic regression \citep{agresti2012categorical}. Additionally, we apply a sensitivity analysis approach proposed by \cite{lin1998assessing} (specifically their (2.18)) to our 1,365 matched sets of broad cases. This sensitivity analysis involves three parameters: the prevalence $P_1$ of a binary unmeasured confounder among the exposed, the prevalence $P_0$ of the binary unmeasured confounder among the unexposed, and the conditional odds ratio $\Xi$ of the binary unmeasured confounder on the outcome conditional on observed covariates. The analysis assumes that the binary unobserved confounder is independent of the observed covariates. In a specified range of values for $P_1$ and $P_0$, satisfying $P_1>P_0$, the worst-case p-value is attained at $P_1=1$ and $P_0=0$, and the worst-case p-value remains significant at $\Xi=3.75$. Applying their analysis to the 1,010 narrow case matched sets, the worst-case p-value is also achieved at $P_1=1$ and $P_0=0$, and the worst-case p-value remains significant at $\Xi=4.77$. Hence, using the narrow case definition still has the potential to  increase insensitivity to unmeasured confounding when using a logistic regression to adjust for observed covariates.  
However,  this narrow-case result should be interpreted with caution because the method in \cite{lin1998assessing} does not account for potential selection bias, a crucial consideration in our study. Developing sensitivity analysis methods for selection bias based on logistic regression is an area for future research.
}

\subsection{Using a Second Referent Group}

{In our primary case-referent study in Section \ref{subsec: application, referent group 1}, the cases were obtained from the 1993 NMFS, while the  living referents were sourced from the 1994 NHIS. Since the cases and referents were from two different sources, there are meaningful differences in data collection between  these two sources \citep{Wiebe:2003aa}. On the one hand, firearm data for the cases were from next of kin who might not know about their firearms. On the other hand, living referents might also disinclined to admit having firearms at home during NHIS survey.  Hence, the direction of bias from the measurement of the firearm information is unclear.

To address this limitation, we conduct another case-referent study using non-injury deaths from the 1993 NMFS as an alternative referent group. The use of multiple referent groups is useful in case-control studies when the referent groups have different potential biases \cite[Chapter 7]{rosenbaum2002}.  If similar results are obtained with each referent group, then the evidence is strengthened in the sense that multiple biases would need to be postulated to explain away the results rather than just one bias.  
Following \citet{grassel2003association}, we exclude homicide and accidental injury deaths from the referent group, because their cause of death may be associated with the exposure of interest (having firearms at home), and among them the exposure may be over-represented, which can lead to underestimating the exposure’s effect \citep{rothman2008modern}.  Using this second referent group, each of the 1,365 broad cases is matched to three similar referents in the same way as described in Section 1.2. The covariate balance of the matched data is in Table \ref{tb1}.



{Results from this analysis are similar to the results in Section \ref{subsec: application, referent group 1}. Therefore, using a second referent group adds an additional piece of evidence that supports our findings between firearms and suicide. Specifically, comparing with and without firearms at home, the odds ratio for suicide is  4.42. Through a randomization test of no effect, a one-sided p-value is  $ 4.90 \times 10^{-126}$. This compelling evidence suggests that the presence of firearms at home increases the risk of suicide if controlling for the observed covariates through successful matching has successfully rendered the matched subjects exchangeable. A sensitivity analysis indicates that the observed effect is insensitive to a bias up to $\Gamma = 4.5$.  Restricting to the narrow case matched sets and comparing with and without firearms at home, the odds ratio for suicide at home is 5.23, and the rejection is insensitive at $(\Theta, \Gamma)=(1.0,5.2)$ or $(\Theta, \Gamma)=(1.22,4.3)$ or $(\Theta, \Gamma)=(1.36,3.8)$.  
Using the combined test, the rejection is insensitive at $(\Theta, \Gamma)=(1.0,5.1)$ or $(\Theta, \Gamma)=(1.22,4.4)$ or $(\Theta, \Gamma)=(1.36,4.4)$. 

We also apply the standard logistic regression approach that adjusts for all observed covariates listed in Table 1. Comparing with and without firearms at home and conditional on all observed covariates, the conditional odds ratio for suicide is 4.97, for suicide at home when restricting to the narrow case matched sets is 6.07. Applying \cite{lin1998assessing}'s sensitivity analysis method to the broad case definition, the worst-case p-value is achieved at $P_1=1$ and $P_0=0$, and the worst-case p-value remains significant at $\Xi=4.31$. Applying their analysis to the 1,010 narrow case matched sets, the worst-case p-value is also achieved at $P_1=1$ and $P_0=0$, and the worst-case p-value remains significant at $\Xi=5.12$, which needs to be interpreted with caution because of not accounting for potential selection bias.}

\section{Discussion}

{In the absence of a gold standard case definition, there is often a broad case definition and a narrow case definition. This is common for event-defined broad cases, as the narrow case can be defined by a specific means of the event happening, cause of the event, place of the event, or severity of the event. For example, the broad case and narrow case pairs could be  (i) all-cause mortality and disease-specific mortality; (ii) disease and late-stage (or high-grade) disease; (iii) disease and disease with a specific subtype; and (iv) suicide and suicide at home (studied in this article).  
The broad case definition offers a larger number of cases, while the narrow case definition often offers a larger effect size because the treatment effect is not diluted by cases for whom the treatment effect is minimal or having misclassified cases. A larger effect size tends to be harder to be explained away by unmeasured confounding bias.  Hence, despite our inference goal being about the broad case, there is a rationale for using the narrow case definition. However, before examining the data, it is uncertain which case definition will provide greater statistical power. This is why we propose a combined test that tests twice using both the broad and narrow case definitions with a correction for multiple testing. In the limit, as the sample size goes to infinity, the combined test has the same insensitivity to unmeasured confounding bias as the better one of the broad case test and the narrow case test. In finite samples, the combined test pays a small price in terms of power compared to knowing a priori which case definition is better, but gains a lot compared to making an incorrect guess about which case definition is better. 

Motivated by our firearms and suicide application, we see that a crucial and delicate issue arises when using the narrow case definition. That is, restricting to narrow cases may introduce selection bias because the treatment can affect case definition. This issue extends beyond our specific context and can also arise in various scenarios.  For example, in studying the effect of cancer screening on all-cause mortality (broad case) and   cancer-specific mortality (narrow case),  cancer screening can lead to more cancer diagnoses, which, in turn, can result in deaths from an uncertain cause being more likely to be attributed to cancer due to a phenomenon called  {sticky-diagnosis bias} \citep{baker2002statistical}. In another example of 
studying the effect of cancer screening on the diagnosis of cancer (broad case) and  the diagnosis of late-stage cancer (narrow case),  cancer screening may  alter the stage at which cancer is diagnosed \citep{chari2022early}. Similarly, when investigating the effect of a treatment on a disease (broad case) and high-grade disease (narrow case), it is common for the treatment to affect the grade of the disease. Our proposed methods use an additional sensitivity parameter to control this selection bias.

In our study of firearms and suicide, we used both suicide (broad case) and suicide at home (narrow case). By conducting a randomization test of Fisher's sharp null hypothesis for the broad case, we obtained strong evidence indicating that the presence of a gun in the home increases the risk of suicide,  if controlling for the observed covariates by matching has successfully rendered the matched subjects exchangeable. This rejection of the null hypothesis is insensitive to unmeasured confounding at $\Gamma=3.9$, which means that one subject in a matched set may have 3.9 times higher odds of having firearms at home than another subject in the matched set. We also applied our proposed  narrow-case test, and we found that the rejections are insensitive at $(\Theta, \Gamma) = (1.22,4.2)$ or $(\Theta, \Gamma) = (1.36,3.8)$. Using the combined test,  the rejections are insensitive at $(\Theta, \Gamma) = (1.22,4.1)$ or $(\Theta, \Gamma) = (1.36,3.8)$. Note that  $ \Theta=1.22 $ means that  among people who were going to commit suicide no matter whether they had a firearm or not,  up to $ 18\%$ of the subjects committing suicide at home when having firearms at home would commit suicide in other places if  without firearms at home; $ \Theta=1.36 $ would allow up to $  26\%$ such subjects. We also provided some informal benchmark that shows a $\Theta=1.36$ is likely able to control the selection bias but may be too conservative. Therefore, whether leveraging the narrow case can strengthen the causal conclusion against unmeasured confounding depends on the magnitude of  $\Theta$. We also conducted another case-referent study using a second referent group and found similar results.

Nonetheless, interpreting the sensitivity parameter $\Theta$ can be challenging in situations where our understanding of the impact of treatment on the case definition is limited. 
Future research requires formal calibration strategies to enhance the interpretability of $\Theta$. Moreover, additional data on important confounders such as mental health problems can help with reducing unmeasured confounding bias and facilitate the calibration of sensitivity parameters. For example, the National Violent Death Reporting System (NVDRS) provides information on mental health problems. However, NVDRS exclusively includes violent deaths, which poses challenges in selecting appropriate referents for comparison \citep{lyons2020selection}. Therefore, linking national surveillance systems and survey data sources with complementary data sources that contain detailed health information, such as electronic health records (EHR)  and claims data, can improve future case-referent studies and suicide research.
}


\vspace{-4mm}

\bibliographystyle{apalike} 
\bibliography{reference}

\spacingset{1.2} 
\begin{table}[ht]
	\centering
	\caption{Covariate balance after matching 1,365  broad cases to two referent groups. \label{tb1}}
	\resizebox{!}{.5\textheight}{\begin{tabular}{lcccccc}
			\hline 
 & &  \multicolumn{2}{c}{Referent Group 1 (from NHIS)} & &  \multicolumn{2}{c}{Referent Group 2 (from NMFS)}  \\    
 \cline{3-4}   \cline{6-7}
			& Broad cases, \%&  
   Referents, \% &  Standardized &&   
   Referents, \% &  Standardized\\ 
			&   ($ I= $1,365)& ($ I= $1,365) & mean difference&& ($ 3I= $4,095) & mean difference \\   \hline
			Firearms in home &  65.8 &  34.4  &  &  & 30.3 \\ 
			   Female &  28.6 &  28.6  & 0.00 &  & 28.3 & 0.01\\ 
			Race  &       &       \\ 
			\quad  White &  88.6  &  88.6 &0.00 & & 88.6 & 0.00\\ 
			\quad  Black &   8.6 &  8.6 &0.00 & & 8.6 & 0.00\\ 
			\quad    Other &   2.9 &  2.9 & 0.00 & & 2.9 & 0.00 \\ 
			Age (mean (SD)) & 48.90 (20.98) & 47.07 (19.46)  & 0.09 & & 50.25 (20.28) & -0.07\\ 
			Lived alone  \\ 
                \quad Yes &  26.5 &  27.3 & -0.02 & & 20.0 & 0.15\\
                \quad No &   71.1 &  72.7 & -0.04 & & 76.9 & -0.13\\ 
                \quad Missing &  2.4 &   0.0 & 0.22 & & 3.1 & -0.04\\
			Marital status&       &       \\ 
			\quad Married &  43.1 &  45.7 &  -0.05 & & 41.8 
 & 0.03\\ 
			\quad Widowed &  13.0 &  12.7 & 0.01 & & 11.9 & 0.03\\ 
			\quad Divorced/separated &   15.3 &  15.0 & 0.01 & & 14.2 & 0.03\\ 
			\quad Never married &   27.7 &  26.0 & 0.04 & & 31.8 & -0.09 \\  
                \quad {Missing} &  0.9 &   0.5 & 0.05 & & 0.2 & 0.03\\ 
			Education &       &       \\ 
			\quad 0-8yrs &       9.1 &   9.0 & 0.00 & & 10.8 & -0.06\\ 
			\quad 9-11yrs &    13.7 &  13.3 & 0.01 & & 12.2 & 0.04 \\ 
			\quad  12yrs &   38.5 &  40.5 & -0.04 & & 39.1 & -0.01\\ 
			\quad 13-15yrs &   15.8 &  17.1 & -0.03 & & 15.6 & 0.01\\ 
			\quad  $ \geq $16yrs & 12.6 &  13.5 & -0.02 & & 11.9 & 0.02\\  
                \quad Missing  &  10.3 &   6.5 & 0.17 & & 10.4 & 0.00 \\ 
			Annual family income &       &       \\ 
			\quad$<$\$1,000 &   2.8 &   2.7 & 0.01 & & 2.4 & 0.02\\ 
			\quad \$1,000- \$1,999 &      0.7 &   0.7 & 0.00 & & 0.4 & 0.03\\ 
			\quad  \$2,000- \$2,999 &   0.6 &   0.6 & 0.00 & & 0.4 & 0.02\\ 
			\quad  \$3,000- \$3,999 &    0.9 &   0.9 & 0.00 & & 0.7 & 0.02\\ 
			\quad  \$4,000- \$4,999 &   1.2 &   1.2 & 0.00 & & 1.5 & -0.02\\ 
			\quad  \$5,000- \$5,999 &   1.6 &   1.5 & 0.01 & & 1.8 & -0.01\\ 
			\quad   \$6,000- \$6,999 &    1.7 &   1.6 & 0.01 & & 1.8 & -0.01\\ 
			\quad  \$7,000- \$8,999 &   4.4 &   4.4 &  0.00 & & 4.8 & -0.02\\ 
			\quad   \$9,000- \$13,999 &   10.0 &   9.5 & 0.02 & & 10.8 & -0.03\\ 
			\quad  \$14,000- \$18,999 &   8.1 &   8.2 & 0.00 & & 8.6 & -0.02\\ 
			\quad   \$19,000- \$24,999 & 8.1 &   8.4 & -0.01 & & 9.4 & -0.05\\ 
			\quad   \$25,000- \$49,999 & 16.9 &  18.2 & -0.03 & & 16.3 & 0.02\\ 
			\quad  $ \geq $\$50,000 & 12.2 &  12.2 &  0.00 & & 9.4 & 0.10\\  
                \quad Missing & 30.8 &  30.0 & 0.02 & & 31.5 & -0.01\\ 
			Veteran  &  \\
                \quad Yes &  26.0 &  22.8 & 0.08 &  &22.6 & 0.08 \\ 
                \quad No &   69.6 &  74.1 & -0.11 &  &73.9 & -0.10\\ 
                \quad Missing &  4.4 &   3.2 &  0.08 & & 3.5 & 0.05\\ 
			Region &       &       \\ 
			\quad Northeast &   11.5 &  12.4 & -0.02 & & 15.4 & -0.11\\ 
			\quad Midwest &   23.6 &  25.8 & -0.05 & & 22.4 & 0.03\\ 
			\quad South &   38.7 &  36.5 & 0.05 & & 36.5 & 0.05\\ 
			\quad West &  26.2 &  25.3 & 0.02 & & 25.7 & 0.01\\ 
			Population  &       &       \\ 
			\quad Non-MSA &  26.9 &  22.7 & 0.10 & & 21.2 & 0.14\\ 
			\quad $<$100,000 &    0.9 &   0.8 & 0.01 & & 0.8 & 0.01\\ 
			\quad100,000-249,999 &   8.0 &   6.9 & 0.04 & & 8.1 & 0.00\\ 
			\quad$ \geq $250,000 & 64.2 &  69.6 & -0.11 & &69.9 & -0.12\\ 
			\hline
	\end{tabular}}
\end{table}

\clearpage

\setcounter{equation}{0}
\setcounter{table}{0}
\setcounter{lemma}{0}
\setcounter{section}{0}
\setcounter{figure}{0}
\setcounter{theorem}{0}
\renewcommand{\theequation}{S\arabic{equation}}
\renewcommand{\thelemma}{S\arabic{lemma}}
\renewcommand{\thetheorem}{S\arabic{theorem}}
\renewcommand{\thefigure}{S\arabic{figure}}
\renewcommand{\thetable}{S\arabic{table}}

\begin{center}
	\spacingset{1.5} 
	\noindent
	{\Large \bf
		Supplementary Material for ``Combining Broad and Narrow Case Definitions in  Matched Case-Control Studies: Firearms in the Home and Suicide Risk''}%
\end{center}

\appendix

\section{Extensions}
In this section, we discuss (i) how to use our method to calculate two-sided p-values; and (ii) how  our results can be extended if instead assuming $ \Theta^{-1} \leq \theta_{Tij}/\theta_{Cij}  \leq \Theta$ with $ \Theta\geq 1 $.  

For part (i), we calculate the two-sided p-value by testing two one-sided hypotheses and using Bonferroni correction. Hence, the two-sided p-value is obtained by computing both one-sided p-values upper bounds, doubling the smaller one, and taking the minimum of this value and 1.  

For part (ii), when instead assuming $ \Theta^{-1} \leq \theta_{Tij}/\theta_{Cij}  \leq \Theta$ with $ \Theta\geq 1 $, we redefine $ \bar p_{ni}   $ as $ \bar p_{ni}  = m_i/\{m_i + (J-m_i)\Theta\Gamma\}$. Then all the results in the main article hold.

\section{Power of Sensitivity Analysis and Design Sensitivity} 
\label{sec: power}

\subsection{General Power Formulas}

For large $ I $ and $ |\mathcal{N}| $, using the normal approximation in  \eqref{eq: bound n},  the one-sided Mantel-Haenszel test using the broad case definition rejects $ H_0 $ at level $ \alpha  $ for all biases in treatment assignment of magnitude at most $ \Gamma $ if $Y_b \geq I  \mu_{\Gamma, b} + \Phi^{-1} (1-\alpha) \sqrt{I}  \sigma_{\Gamma, b}$. 
The one-sided Mantel-Haenszel test 
using the narrow case definition rejects $ H_0 $ at level $ \alpha  $ for all biases in treatment assignment of magnitude at most $ \Gamma $ and all  effects of the treatment on case definition among the always-cases of magnitude at most $ \Theta $ if $Y_n \geq 	|\mathcal{N}|  \mu_{\Gamma\Theta, n} + \Phi^{-1} (1-\alpha) \sqrt{	|\mathcal{N}|}  \sigma_{\Gamma\Theta, n},$
where $ \Phi^{-1} (1-\alpha) $ is the $ 1-\alpha $ quantile of the standard normal distribution,  and 
\begin{align*}
	&\mu_{\Gamma,b}=\frac{  1 }{I} \sum_{i=1}^I \bar{\bar p}_{bi},\quad  \sigma^2_{\Gamma,b} = \frac{1}{I} \sum_{i=1}^I \bar{\bar p}_{bi} (1-  \bar{\bar p}_{bi}), \\
	&  \mu_{\Gamma\Theta,n}= \frac{1}{	|\mathcal{N}| }	 \sum_{i\in \mathcal{N}}  \bar{\bar p}_{ni},  \quad   \sigma^2_{\Gamma\Theta,n} = \frac{1}{|\mathcal{N}|}  \sum_{i\in \mathcal{N}} \bar{\bar p}_{ni}(1-\bar{\bar p}_{ni}).
\end{align*}

Next, we determine the power of sensitivity analysis and design sensitivity based on $ Y_b $ and $ Y_n $  \citep[Chapter 14]{rosenbaum2004design, rosenbaum2010design}. The power of a sensitivity analysis is the probability, for given values of $ \Gamma $ and $ \Theta $ of the sensitivity parameters and a given significance level $ \alpha $, that Fisher's $ H_0 $ will be rejected under the favorable situation when there is a treatment effect and no unmeasured confounding.  As $ I $ and $ |\mathcal{N}| $ go to infinity, for a given $ \Theta $,  there is generally a number $ \tilde \Gamma $, called the design sensitivity, such that the power tends to 1 if $ \Gamma<\tilde{\Gamma} $ and to 0 if $ \Gamma>\tilde{\Gamma} $. Among other things, the design sensitivity $ \tilde\Gamma $  characterizes the limiting behavior of a test and is a guide to choosing tests that are more robust to unmeasured biases; see,  for instance, \citet{Stuart:2013aa}  and \citet{Zubizarreta:2013aa} for related discussions in cohort studies.

Given the design of matched case-referent studies, it is quite natural to conduct the power analysis of $ Y_b $ conditioning on the event  $ \mathcal{S} = \{\kappa_b(R_{i1})= 1, \kappa_b(R_{i2})= \dots = \kappa_b(R_{iJ})= 0,  i=1,\dots, I\} $, i.e., each matched set is composed of one broad case (indexed by  subject $ i1 $) and $ J-1 $ referents. As such, the probabilistic statement is made with respect to  a theoretical sequence of case-referent studies with $ I $ broad case matched sets while allowing a random subset of them to be the narrow case matched sets.	To compare the power of $ Y_b $ and $ Y_n $ based on the same frame of reference, the power analysis of $ Y_n $ also conditions on the event $ \mathcal{S} $. {The technical details used to establish the power results are adapted from the literature of sensitivity analysis; see e.g., \cite{Small:2013aa}.}

By Lindeberg's central limit theorem, the test statistics $ Y_b $ and $Y_n $ are asymptotically normal as $ I $ and $ |\mathcal{N}| $ go to infinity,  that is
\begin{align}
	&\frac{\sqrt{I}(Y_b/I- \mu_{b} ) }{ \sigma_{ b}}  \mid \mathcal{S} \xrightarrow{d} N(0,1) \text{ and } \frac{\sqrt{	|\mathcal{N}|} (Y_n/	|\mathcal{N}|- \mu_{n})}{ \sigma_{n}}  \mid \mathcal{S}, \mathcal{N}  \xrightarrow{d} N(0,1), \nonumber
\end{align}
where  $ \mu_b=I^{-1} \sum_{i=1}^{I} p_{bi} $, $ \mu_n=|\mathcal{N}|^{-1} \sum_{i\in \mathcal{N}} p_{ni} $, $ p_{bi} = \Pr(Z_{i1}= 1\mid \kappa_b(R_{i1} )= 1) $, $ p_{ni} = \Pr(Z_{i1}= 1\mid \kappa_b(R_{i1} ) =\kappa_n(R_{i1} )  = 1) $, provided that the two limits  $ \lim_{I\rightarrow\infty}  I^{-1} \sum_{i=1}^I  p_{bi} (1-p_{bi}) = \sigma_{b}^2 >0 $ and $ \lim_{|\mathcal{N}|\rightarrow\infty} |\mathcal{N}|^{-1} \sum_{i\in \mathcal{N}}  p_{ni} (1-p_{ni}) = \sigma_{n}^2  >0 $ exist. From the dominated convergence theorem, the convergence result for $ Y_n $ also holds when conditioning on $ \mathcal{S} $.   After some algebra, the power of $ Y_b $ with the sensitivity parameter $ \Gamma $ is given by  
\begin{align}
	\pr \left( Y_b \geq I \mu_{\Gamma, b} +\Phi^{-1}(1-\alpha) \sqrt{I} \sigma_{\Gamma, b}  \mid \mathcal{S}
	\right) \approx  \Phi \left( \frac{\sqrt{I} (\mu_{b} - \mu_{\Gamma, b}) -\Phi^{-1}(1-\alpha)   \sigma_{\Gamma, b}  }{\sigma_{b}}\right). \label{eq: power1}
\end{align}
Using $ q I$  to approximate  $|\mathcal{N}| $, where  $q = I^{-1} \sum_{i=1}^{I} \pr (\kappa_n (R_{i1})= 1\mid \kappa_b(R_{i1})= 1) > 0$,  the power of $ Y_n $ with the sensitivity parameters $ \Gamma $  and $ \Theta $ is given by  
\begin{align}
	&\pr \left( Y_n \geq |\mathcal{N}| \mu_{\Gamma\Theta, n} +  \Phi^{-1}(1-\alpha)  |\mathcal{N}|^{1/2}\sigma_{\Gamma\Theta, n} 
	\mid \mathcal{S} 	\right)\approx \Phi \left( \frac{\sqrt{qI} (\mu_{ n} - \mu_{\Gamma\Theta, n}) - \Phi^{-1}(1-\alpha) \sigma_{\Gamma\Theta, n}  }{\sigma_{n}}\right).\label{eq: power2} 
\end{align}

\subsection{An Example}
\label{subsec: eg}
The power formulas \eqref{eq: power1}-\eqref{eq: power2} are quite general but are  abstract. In this section, we provide a simple but important example to illustrate in detail the power of sensitivity analysis and design sensitivity.

Consider a favorable situation when there is a treatment effect and no unmeasured confounding. Suppose that $ (Z_{ij}, \kappa_b(R_{ij} ), \kappa_n(R_{ij} ) ) $, $ i=1,\dots, I, j=1,\dots, J $ are independent and identically distributed, satisfying  $ Z_{ij} \perp (\kappa_b(r_{Tij}), \kappa_b(r_{Cij}) ,  \kappa_n(r_{Tij }), \kappa_n(r_{Cij })) $ and 
\begin{equation} \label{eq: favorable}
	\begin{aligned}
		&\Pr(Z_{ij}= 1) = \pi, \\
		&\Pr(\kappa_b (R_{ij})= 1\mid Z_{ij}= 1)= b_T, \\
		&\Pr(\kappa_b (R_{ij})= 1\mid Z_{ij}= 0) = b_C, \\ 
		&\Pr(\kappa_n (R_{ij})= 1\mid\kappa_b (R_{ij})= 1, Z_{ij}= 1) = \eta_T ,\\
		&\Pr(\kappa_n (R_{ij})= 1\mid \kappa_b (R_{ij})= 1, Z_{ij}= 0)= \eta_C .
	\end{aligned}
\end{equation}
Note that technically it is easy to accommodate measured covariates by specifying $ b_T, b_C, \eta_T, \eta_C $ as functions of observed covariates, but this complicates our illustration. 

Imagine the following data generating process. First, we obtain $ I$ broad cases and $ I(J-1) $ referents. Renumber the subjects such that the first subject in each matched set is the broad case, the rest are the referents, i.e.,  for every $ i=1,\dots, I $, $ \kappa_b(R_{i1}) = 1 $ and $ \kappa_b(R_{i2}) = \dots= \kappa_b(R_{iJ}) = 0 $. Second,  given the broad case variables, we generate the exposure variables $ (Z_{ij}, i=1,\dots,I, j=1,\dots, J) $  according to  
\begin{align}
	&\Pr(Z_{ij}= 1\mid \kappa_b(R_{ij} )= 1) = \frac{b_T\pi}{b_T\pi + b_C (1-\pi)},\label{eq: pbi}\\ 
	&\Pr(Z_{ij}= 1\mid \kappa_b(R_{ij} )= 0) = \frac{(1-b_T)\pi}{(1-b_T)\pi + (1-b_C) (1-\pi)}, \nonumber
\end{align} 
which are derived from  \eqref{eq: favorable} and  the Bayes rule. 
Third, given the broad case variables and the exposure variables, we  generate  the narrow case variables  $ (\kappa_n(R_{ij}), i=1,\dots,I, j=1,\dots, J) $ according to \eqref{eq: favorable}. It is important to realize that this data generating process conditions on $ \mathcal{S} $ and is used as the frame of reference for the power analysis.

Under the setting described in \eqref{eq: favorable}, by the weak law of large numbers, we have that $ \mu_{\Gamma, b}= I^{-1} \sum_{i=1}^I \bar{\bar p}_{bi} \mid \mathcal{S}\xrightarrow{p} \E( \bar{\bar p}_{bi} \mid \mathcal{S}) $ and $\sigma_{\Gamma, b}^2=  I^{-1} \sum_{i=1}^I \bar{\bar p}_{bi} (1-  \bar{\bar p}_{bi}) \mid \mathcal{S}\xrightarrow{p}   \E\left\{\bar{\bar p}_{bi} (1-\bar{\bar p}_{bi}) \mid \mathcal{S}\right\}  $. Hence, the power of sensitivity analysis based on $ Y_b $ in  \eqref{eq: power1} can be simplified as 
\begin{align}
	&\pr \left( Y_b \geq I \mu_{\Gamma, b} +\Phi^{-1}(1-\alpha) \sqrt{I} \sigma_{\Gamma, b}  \mid \mathcal{S}
	\right) \label{eq: power b} \\
	&\qquad \approx 	\Phi\left( \frac{\sqrt{I} \left\{p_{bi}    -\E(\bar{\bar p}_{bi}\mid \mathcal{S}) \right\}- \Phi^{-1} (1-\alpha) \sqrt{  \E\left\{\bar{\bar p}_{bi} (1-\bar{\bar p}_{bi}) \mid \mathcal{S}\right\}  } } {\sqrt{p_{bi} (1-p_{bi})  }} \right), \nonumber
\end{align}
where  $ p_{bi} = \Pr(Z_{i1}= 1\mid \kappa_b(R_{i1} )= 1) $ is expressed in \eqref{eq: pbi}, and $  \bar{\bar p}_{bi}$ is defined in \eqref{eq: SA broad}. In Section \ref{subsec: additional details}, we give the explicit expression of the probability mass function of $ m_i $ conditional on $ \mathcal{S} $, and thus the explicit expressions of $ \E (\bar{\bar p}_{bi} \mid \mathcal{S}) $ and $ \E\{\bar{\bar p}_{bi} (1- \bar{\bar p}_{bi}) \mid \mathcal{S}\} $.  As $ I\rightarrow \infty$, the power of design sensitivity with the given $ \Gamma $ is driven by the sign of $ p_{bi} -\E(\bar{\bar p}_{bi} \mid \mathcal{S}) $: if $p_{bi}    -\E(\bar{\bar p}_{bi} \mid \mathcal{S}) <0 $, the power tends to 0, while if $  p_{bi}-\E(\bar{\bar p}_{bi} \mid \mathcal{S}) >0 $, the power tends to 1. By definition, the design sensitivity using the broad case definition $ \tilde{\Gamma}_b $ is the solution in $ \Gamma $ to the equation $  p_{bi} -\E(\bar{\bar p}_{bi} \mid \mathcal{S}) =0$, which as shown in Section \ref{subsec: additional details}, has a simple analytical form
\begin{align}
	\tilde{\Gamma}_b = \frac{b_T/(1-b_T)}{b_C/(1-b_C)}. \nonumber
\end{align}

For the narrow case definition, from the facts that  $q= I^{-1} \sum_{i=1}^{I} \Pr (\kappa_n (R_{i1})= 1\mid \kappa_b(R_{i1})= 1) = \{\eta_T b_T\pi + \eta_Cb_C(1-\pi) \} / \{ b_T\pi + b_C(1-\pi)\}$, 
$ \mu_{\Gamma\Theta, n}= |\mathcal{N}|^{-1} \sum_{i\in \mathcal{N}} \bar{\bar p}_{ni} \mid \mathcal{S}, \mathcal{N}\xrightarrow{p} \E( \bar{\bar p}_{ni} \mid \mathcal{S}, i\in \mathcal{N}) $, and $\sigma_{\Gamma\Theta, n}^2=  |\mathcal{N}|^{-1} \sum_{i\in \mathcal{N}} \bar{\bar p}_{ni} (1-  \bar{\bar p}_{bi}) \mid \mathcal{S}, \mathcal{N}\xrightarrow{p}   \E\left\{\bar{\bar p}_{ni} (1-\bar{\bar p}_{ni}) \mid \mathcal{S}, i\in \mathcal{N}\right\}  $,  
the power of sensitivity analysis based on $ Y_n $ in \eqref{eq: power2} can be simplified as 
\begin{align}
	&	\pr \left( Y_n \geq |\mathcal{N}| \mu_{\Gamma\Theta, n} +  \Phi^{-1}(1-\alpha)  |\mathcal{N}|^{1/2}\sigma_{\Gamma\Theta, n} 
	\mid \mathcal{S}	\right) \label{eq: power n} \\
	&\qquad \approx
	\Phi\left( \frac{\sqrt{qI } \left\{ p_{ni}  - \E( \bar{\bar p}_{ni}\mid \mathcal{S},i\in \mathcal{N}) \right\} -  \Phi^{-1} (1-\alpha)\sqrt{  \E\left\{\bar{\bar p}_{ni} (1-\bar{\bar p}_{ni})\mid \mathcal{S},i\in \mathcal{N}\right\}  } } { \sqrt{ p_{ni} (1-p_{ni}) }} \right), \nonumber 
\end{align}
where $ p_{ni} =\pr(Z_{i1} = 1\mid \kappa_b(R_{i1})= \kappa_n(R_{i1})= 1) = b_T \eta_T \pi/ \{ b_T\eta_T \pi + b_C\eta_C(1-\pi)\}$ from the Bayes rule, and  $  \bar{\bar p}_{ni}$ is defined in \eqref{eq: prob bound}. Again, we give the explicit expressions of $  \E( \bar{\bar p}_{ni}\mid \mathcal{S},i\in \mathcal{N}) $ and $   \E\left\{\bar{\bar p}_{ni} (1-\bar{\bar p}_{ni})\mid \mathcal{S},i\in \mathcal{N}\right\} $ in Section \ref{subsec: additional details}.  As $ I \rightarrow \infty$, with the given $ \Theta $, the design sensitivity using the narrow case definition also has a simple analytical form
\begin{align}
	\tilde{\Gamma}_n = \frac{b_T/(1-b_T)}{b_C/(1-b_C)} \times \frac{\eta_T/\eta_C}{\Theta} \nonumber. 
\end{align}
In the next subsection (Section \ref{subsec: some power calculation}), we compare the calculated power using \eqref{eq: power b}-\eqref{eq: power n} and the  power obtained by simulation, finding good agreement.

When can we use the narrow case definition to increase design sensitivity? Based on this simple example, the answer is when $ \eta_T/\eta_C\geq \Theta $, that is when 
\begin{align*}
	\frac{	\Pr(\kappa_n (R_{ij})= 1\mid\kappa_b (R_{ij})= 1, Z_{ij}= 1) }{ \Pr(\kappa_n (R_{ij})= 1\mid\kappa_b (R_{ij})= 1, Z_{ij}= 0) } \geq   \frac{\pr(\kappa_n(R_{ij})=1\mid \kappa_b(r_{Cij})=  \kappa_b(r_{Tij})= 1,  Z_{ij} = 1)}{\pr(\kappa_n(R_{ij})=1\mid  \kappa_b(r_{Cij})=  \kappa_b(r_{Tij})= 1, Z_{ij}=0 )}. 
\end{align*}
{In the context of our application, the left-hand side is the probability ratio of a broad case being a narrow case with and without the firearm. The right-hand side is the same quantity evaluated among the always-cases.  Suppose that case monotonicity holds (i.e., $ \kappa_n(r_{Tij}) \geq  \kappa_n(r_{Cij})$ and $ \kappa_b(r_{Tij}) \geq  \kappa_b(r_{Cij})$ for every $ i,j $), then $ \eta_T/\eta_C\geq \Theta $ when}
\begin{align}
	\pr (\kappa_n (r_{Tij}) = 1\mid \kappa_b (r_{Tij})= 1,  \kappa_b (r_{Cij})= 0) \geq 	\pr (\kappa_n (r_{Tij}) = 1\mid \kappa_b (r_{Tij})= 1,  \kappa_b (r_{Cij})= 1) \label{eq: increase power}.
\end{align}


We can use Figure \ref{fig: 2} to understand \eqref{eq: increase power}. In Figure \ref{fig: 2}, the left panel is the case allocation for all the subjects when untreated (no firearms at home), the right panel is the case allocation for all the subjects when treated (with firearms at home). Figure \ref{fig: 2}  shows that in expectation, $ h_{NN} $ narrow cases, $ h_{MM} $ marginal cases, and $ h_{RR} $ referents are not affected by the treatment. Assuming the case monotonicity, the treatment moves $ h_{MN} $ marginal cases to be narrow cases, $ h_{RN} $ referents to be narrow cases, and $ h_{RM} $ referents to be marginal cases. Then \eqref{eq: increase power} holds when 
\begin{align*}
	\frac{h_{NN}+h_{MN}}{h_{NN}+h_{MN}+h_{MM}}\leq 	\frac{h_{RN}}{h_{RN}+h_{RM}} .
\end{align*}

\begin{figure}[t]
	\centering
	\resizebox{0.8\textwidth}{!}{
		\begin{tikzpicture}
			\draw[draw=black] (10,5) rectangle ++(6,1.5);
			\draw[draw=black] (17,5) rectangle ++(6,1.5);
			\draw[draw=black] (10,3) rectangle ++(6,1.5);
			\draw[draw=black] (17,3) rectangle ++(6,1.5);
			\draw[draw=black] (10,1) rectangle ++(6,1.5);
			\draw[draw=black] (17,1) rectangle ++(6,1.5);
			\draw[draw=black] (17,3) rectangle ++(6,1.5);
			\node[above] at (13,7) {No firearms at home};
			\node[above] at (20,7) {Have firearms at home};
			\node[right, align=left] at (5.5,5.5) {Suicide at home\\ (narrow case)};
			\node[right, align=left] at (5.5,3.5) {Suicide not at home \\ (marginal case)};
			\node[right, align=left] at (5.5,1.5) {Not committing suicide\\ (referent)};
			\node[right] at (10.2,3.8) {$ \{ij\mid \kappa_b(r_{Cij})= 1, \kappa_n(r_{Cij})= 0 \} $ };
			\node[right] at (17.2,3.8) {$ \{ij\mid \kappa_b(r_{Tij})= 1, \kappa_n(r_{Tij})= 0 \} $ };
			\node[right] at (11.4,5.8) {$ \{ij\mid \kappa_n(r_{Cij})= 1 \} $ };
			\node[right] at (18.4,5.8) {$ \{ij\mid \kappa_n(r_{Tij})= 1 \} $ };
			\node[right] at (11.4,1.8) {$ \{ij\mid \kappa_b(r_{Cij})= 0 \} $ };
			\node[right] at (18.4,1.8) {$ \{ij\mid \kappa_b(r_{Tij})= 0 \} $ };
			\draw[->,dblue] (15,4.2) -- (18,5.5) node[pos=1,right] {$ h_{MN} $};
			\draw[->,dblue] (15,6.2) -- (18,6.2) node[pos=1,right] {$ h_{NN} $};
			\draw[->,dblue] (15,3.4) -- (18,3.4) node[pos=1,right] {$ h_{MM} $};
			\draw[->,dorange] (15,1.3) -- (18,1.3) node[pos=1,right] {$ h_{RR} $};
			\draw[->,dorange] (15,2.2) -- (18,3.1) node[pos=1,right] {$ h_{RM} $};
			\draw[->,dorange] (15,2.2) -- (18,5.2) node[pos=1,right] {$ h_{RN} $};
	\end{tikzpicture}}
	\caption{An illustration of how the cases are moved by the treatment under case monotonicity ($  \kappa_n(r_{Tij}) \geq \kappa_n(r_{Cij})$ and $  \kappa_b(r_{Tij}) \geq \kappa_b(r_{Cij})$  for all $ i,j $) in the context of our application.   The left and right panels are respectively the case allocation when all the subjects
		are untreated and when all the subjects
		are treated. The numbers $ h_{NN}, h_{MN}, h_{MM}, h_{RN}, h_{RM},h_{RR} $ are expected values.  \label{fig: 2}}
\end{figure}

The above power analysis results are quite intuitive. With all the other parameters fixed,  if $ \Theta $ is small, i.e., the bias due to restricting to the narrow cases matched sets  is small, then the narrow case test will likely have higher power. On the other hand, if $ \Theta $  is large, i.e., the bias due to restricting to the narrow cases matched sets  is large, then the broad case test will likely have higher power. In many applications, it would be difficult to know this before examining the data. It is in these situations that using the Bonferroni method to combine the broad and narrow case definitions is especially attractive, because the combined procedure has the larger of the two design sensitivities for the two individual tests. It is easy to see this fact. The combined test uses the smaller of the two p-values from the individual tests multiplied by 2. For any fixed $ \Theta $, if the sensitivity parameter $ \Gamma $ is below $ \tilde\Gamma_b $ and $ \tilde\Gamma_n $, then the smaller of the two p-values will tend to zero as $  I \rightarrow \infty$ and thus the p-value from the combined test will tend to zero. This justifies that the combined test will have power approaching one to reject $ H_0 $ for $ \Gamma<\max(\tilde{\Gamma}_b, \tilde{\Gamma}_n) $. For the other direction, if $ \Gamma>  \max(\tilde{\Gamma}_b, \tilde{\Gamma}_n)$, then both p-values will tend to one as $I\rightarrow \infty $ so the combined test  will also have power approaching zero.

\subsection{Some Power Calculations}
\label{subsec: some power calculation}

Figure \ref{fig: supp sim} depicts the calculated power using \eqref{eq: power b}-\eqref{eq: power n} for the broad and narrow case tests under the  data generating process described in Section \ref{subsec: eg}, with $  \alpha=0.05, J=6, \pi =1/3, b_T=0.3, b_C=0.1, \eta_T=0.3, \eta_C=0.15$. 
Figure \ref{fig: supp sim} also shows simulation powers, which are obtained from counting the  proportions of rejections. In general, we find good agreement comparing the powers by formula and powers by simulation. The power of the broad case test does not depend on $ \Theta $, so the broad case power curves for different values of $ \Theta $ are the same. Because the design sensitivity of the broad case test is $ \tilde\Gamma_b= 3.86  $,  the power of the broad case test increases to 1 for $ \Gamma<3.86= \tilde\Gamma_b$ and decreases to zero for  $ \Gamma>3.86= \tilde\Gamma_b$. Analogously,  the design sensitivity of the narrow case test is  $ \tilde\Gamma_n=7.71/\Theta$, the power of the narrow case test increases to 1 for $ \Gamma  < 7.71/\Theta= \tilde\Gamma_n$ and decreases to zero for  $ \Gamma  > 7.71/\Theta= \tilde\Gamma_n$.

Figure \ref{fig: supp sim}  also shows the simulated  power of the combined test; that is, the percentage of multiplicity-adjusted upper bounds on p-values falling below 0.05 in 3,000 simulation runs. Because the combined test attains the maximum design sensitivity of the broad and narrow case tests, the power of the combined test increases to 1 when there is one component test whose power increases to 1.

From Figure \ref{fig: supp sim}, we see that when the narrow case definition offers a considerably larger effect size,  the narrow case test has superior or at least similar power performance when the adjustment of selection bias is small, e.g., when $ \Theta=1 $ or  $ 1.5 $. Note that the x-axis is the number of broad case matched sets. This means that in a conventional case-referent study where cases and referents would be interviewed to collect treatment and covariate information, the costs of the narrow case study would be smaller, but its conclusion would be more robust to unmeasured confounding. However, this superiority of the narrow case test vanishes when the adjustment of selection bias grows, e.g., when $ \Theta=2 $.  In practice, when the investigator is unsure about which case definition gives a conclusion that is more robust to unmeasured confounding, the combined test that uses both case definitions usually has good  performance.

Table \ref{tb: supp simu} considers a wider variety of scenarios and reports the design sensitivity and  simulation power for the given $ I $, where $ I $ is calculated using \eqref{eq: power b} for the broad case test to achieve 80\% power when $ b_C=0.10, b_T=0.30 $. The powers calculated by formulas \eqref{eq: power b}-\eqref{eq: power n} are similar to the simulated powers; see the supplementary materials. 

Overall, there are four takeaways from this simulation study. First,  leveraging the narrow case definition may not be very helpful in the primary analysis with $ \Gamma=1 $ because the broad case definition offers a larger sample size. That's why under scenarios in which $ \tilde{\Gamma}_n>\tilde{\Gamma}_b $, the broad case test is still seen to be more powerful. Second, although
we are reassured that the narrow case test has larger design sensitivity than the broad case test when $ \eta_T/\eta_C>\Theta $, this does not necessarily mean that the sensitivity analysis based on the narrow case definition is more powerful than that based on the broad case definition (e.g., see the fifth line when $ \Gamma=3, \Theta=1 $). Under these circumstances, the combined test that uses both case definitions usually has good performance, or even larger power compared to each component test. Third, when the adjustment for selection bias is large, i.e., when $\Theta $ exceeds $ \eta_T/\eta_C $, leveraging the narrow case definition no longer helps. Finally, the situation with $ I=3785 $ is close to what is predicted by the design sensitivity.

\begin{figure}[h]
	\includegraphics[width=\linewidth]{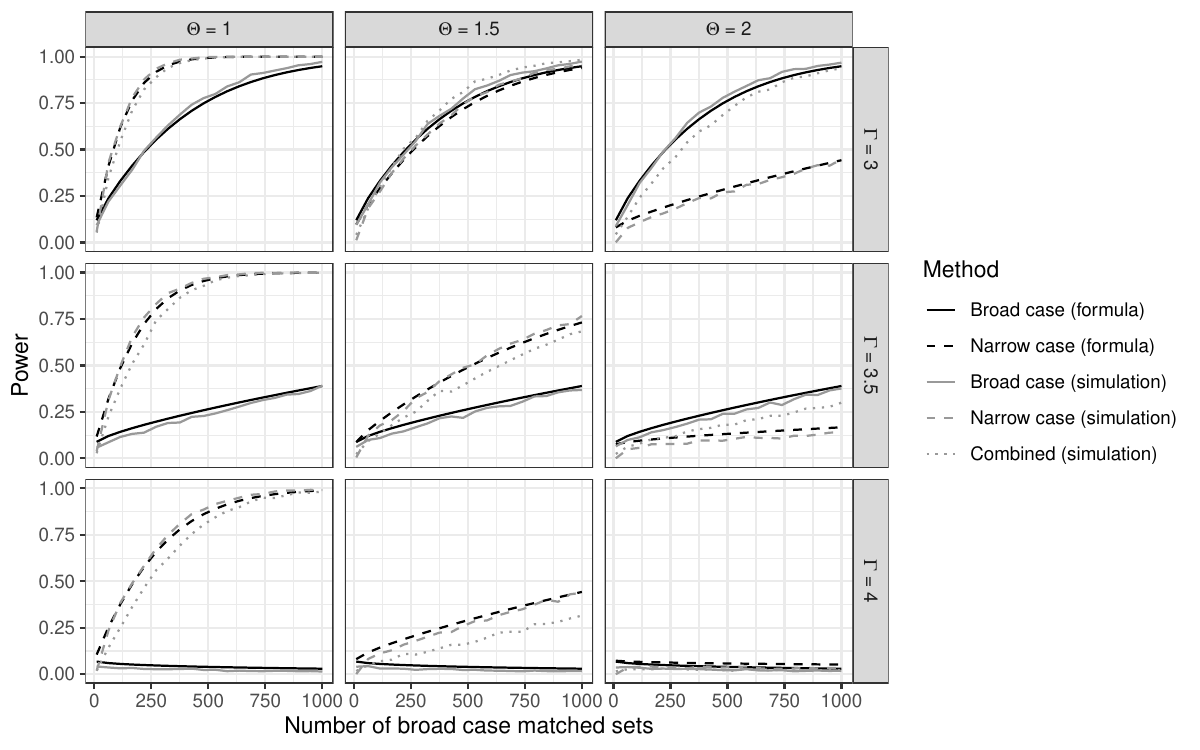}
	\caption{Power of sensitivity analysis for the broad and narrow case tests, where the x-axis represents the number of broad case matched sets $ I $, the simulation powers are obtained using 3,000 repetitions. 
		The data generating process is described in Section \ref{subsec: eg}, with $ J=6, \pi=1/3,  b_T=0.3, b_C=0.1, \eta_T=0.3, \eta_C=0.15.$ The design sensitivity for the broad case test is $ \tilde\Gamma_b= 3.86 $, for the narrow case test is $ \tilde\Gamma_n= 7.71/\Theta $. 
		\label{fig: supp sim}}
\end{figure}

\begin{table}[ht]
	\centering
	\caption{Design sensitivity and simulated power (based on 3,000 simulation runs) for 0.05-level sensitivity analyses using the broad case, narrow case, and both case definitions combined using the Bonferroni method; $\E |\mathcal{N}| $ is the expected number of narrow case matched sets. 
		The data generating process is described in Section \ref{subsec: eg}, with $ J=6 $  and  $\pi=1/3 $. \label{tb: supp simu}} \vspace{-4.2mm}
	\resizebox{.8\textwidth}{!}{\begin{tabular}{cccccccccccccccccc} \hline \\[-2.5ex]
			&&&&&&&&& &		 && \multicolumn{2}{c}{Formula power,  \%} &&  \multicolumn{3}{c}{Simulated power,  \%} \\ 
			$ \Gamma $ & $ \Theta $ & $ I $ &$ b_C $   & $ b_T $   & $ \eta_C $   & $ \eta_T $   &&$\E |\mathcal{N}| $ & $ \tilde{\Gamma}_b $ & $ \tilde{\Gamma}_n $&	&Broad case&  Narrow case && Broad case&  Narrow case & Combined \\\hline \\[-2.5ex]
			1   & 1   & 18   & 0.01 & 0.03 & 0.80 & 0.85 &  & 15   & 3.1 & 3.3 &  & 68.0 & 65.1  &  & 70.5 & 66.7  & 64.0  \\
			&     &      & 0.01 & 0.03 & 0.15 & 0.20 &  & 3    & 3.1 & 4.1 &  & 68.0 & 31.4  &  & 69.2 & 31.2  & 61.0  \\
			&     &      & 0.01 & 0.03 & 0.15 & 0.30 &  & 4    & 3.1 & 6.1 &  & 68.0 & 48.8  &  & 70.7 & 50.9  & 66.3  \\
			&     &      & 0.10 & 0.30 & 0.80 & 0.85 &  & 15   & 3.9 & 4.1 &  & 80.4 & 77.2  &  & 82.8 & 79.7  & 78.2  \\
			&     &      & 0.10 & 0.30 & 0.15 & 0.20 &  & 3    & 3.9 & 5.1 &  & 80.4 & 38.3  &  & 83.7 & 37.1  & 74.7  \\
			&     &      & 0.10 & 0.30 & 0.15 & 0.30 &  & 4    & 3.9 & 7.7 &  & 80.4 & 57.7  &  & 82.2 & 58.0  & 79.2  \\
			&     &      &      &      &      &      &  &      &     &     &  &      &       &  &      &       &       \\
			1   & 1.5 & 18   & 0.01 & 0.03 & 0.80 & 0.85 &  & 15   & 3.1 & 2.2 &  & 68.0 & 38.4  &  & 69.7 & 38.1  & 58.9  \\
			&     &      & 0.01 & 0.03 & 0.15 & 0.20 &  & 3    & 3.1 & 2.7 &  & 68.0 & 20.3  &  & 69.3 & 19.0  & 58.5  \\
			&     &      & 0.01 & 0.03 & 0.15 & 0.30 &  & 4    & 3.1 & 4.1 &  & 68.0 & 32.2  &  & 69.3 & 31.4  & 61.0  \\
			&     &      & 0.10 & 0.30 & 0.80 & 0.85 &  & 15   & 3.9 & 2.7 &  & 80.4 & 52.7  &  & 83.5 & 56.0  & 75.4  \\
			&     &      & 0.10 & 0.30 & 0.15 & 0.20 &  & 3    & 3.9 & 3.4 &  & 80.4 & 25.7  &  & 82.8 & 24.5  & 73.4  \\
			&     &      & 0.10 & 0.30 & 0.15 & 0.30 &  & 4    & 3.9 & 5.1 &  & 80.4 & 40.2  &  & 82.9 & 42.0  & 76.9  \\
			&     &      &      &      &      &      &  &      &     &     &  &      &       &  &      &       &       \\
			1   & 2   & 18   & 0.01 & 0.03 & 0.80 & 0.85 &  & 15   & 3.1 & 1.6 &  & 68.0 & 22.3  &  & 71.7 & 20.8  & 58.8  \\
			&     &      & 0.01 & 0.03 & 0.15 & 0.20 &  & 3    & 3.1 & 2.0 &  & 68.0 & 14.6  &  & 70.8 & 9.7   & 58.4  \\
			&     &      & 0.01 & 0.03 & 0.15 & 0.30 &  & 4    & 3.1 & 3.1 &  & 68.0 & 23.0  &  & 70.3 & 20.1  & 58.5  \\
			&     &      & 0.10 & 0.30 & 0.80 & 0.85 &  & 15   & 3.9 & 2.0 &  & 80.4 & 34.3  &  & 82.0 & 32.6  & 73.3  \\
			&     &      & 0.10 & 0.30 & 0.15 & 0.20 &  & 3    & 3.9 & 2.6 &  & 80.4 & 18.9  &  & 82.4 & 14.6  & 71.7  \\
			&     &      & 0.10 & 0.30 & 0.15 & 0.30 &  & 4    & 3.9 & 3.9 &  & 80.4 & 29.5  &  & 83.2 & 29.0  & 75.4  \\
			&     &      &      &      &      &      &  &      &     &     &  &      &       &  &      &       &       \\
			3   & 1   & 559  & 0.01 & 0.03 & 0.80 & 0.85 &  & 464  & 3.1 & 3.3 &  & 9.9  & 21.2  &  & 7.7  & 19.4  & 11.9  \\
			&     &      & 0.01 & 0.03 & 0.15 & 0.20 &  & 101  & 3.1 & 4.1 &  & 9.9  & 37.7  &  & 7.2  & 37.3  & 27.5  \\
			&     &      & 0.01 & 0.03 & 0.15 & 0.30 &  & 134  & 3.1 & 6.1 &  & 9.9  & 95.1  &  & 8.1  & 96.6  & 92.9  \\
			&     &      & 0.10 & 0.30 & 0.80 & 0.85 &  & 464  & 3.9 & 4.1 &  & 80.0 & 87.3  &  & 82.7 & 89.9  & 84.5  \\
			&     &      & 0.10 & 0.30 & 0.15 & 0.20 &  & 101  & 3.9 & 5.1 &  & 80.0 & 72.2  &  & 82.7 & 76.5  & 85.7  \\
			&     &      & 0.10 & 0.30 & 0.15 & 0.30 &  & 134  & 3.9 & 7.7 &  & 80.0 & 99.7  &  & 83.0 & 99.8  & 99.6  \\
			&     &      &      &      &      &      &  &      &     &     &  &      &       &  &      &       &       \\
			3   & 1.5 & 559  & 0.01 & 0.03 & 0.80 & 0.85 &  & 464  & 3.1 & 2.2 &  & 9.9  & 0.0   &  & 7.9  & 0.0   & 4.4   \\
			&     &      & 0.01 & 0.03 & 0.15 & 0.20 &  & 101  & 3.1 & 2.7 &  & 9.9  & 3.5   &  & 8.6  & 1.7   & 5.2   \\
			&     &      & 0.01 & 0.03 & 0.15 & 0.30 &  & 134  & 3.1 & 4.1 &  & 9.9  & 40.5  &  & 7.2  & 39.2  & 28.3  \\
			&     &      & 0.10 & 0.30 & 0.80 & 0.85 &  & 464  & 3.9 & 2.7 &  & 80.0 & 1.4   &  & 83.9 & 0.6   & 74.4  \\
			&     &      & 0.10 & 0.30 & 0.15 & 0.20 &  & 101  & 3.9 & 3.4 &  & 80.0 & 16.4  &  & 81.5 & 13.1  & 73.2  \\
			&     &      & 0.10 & 0.30 & 0.15 & 0.30 &  & 134  & 3.9 & 5.1 &  & 80.0 & 77.3  &  & 83.1 & 79.7  & 86.8  \\
			&     &      &      &      &      &      &  &      &     &     &  &      &       &  &      &       &       \\
			3   & 2   & 559  & 0.01 & 0.03 & 0.80 & 0.85 &  & 464  & 3.1 & 1.6 &  & 9.9  & 0.0   &  & 6.8  & 0.0   & 3.6   \\
			&     &      & 0.01 & 0.03 & 0.15 & 0.20 &  & 101  & 3.1 & 2.0 &  & 9.9  & 0.3   &  & 7.2  & 0.1   & 3.5   \\
			&     &      & 0.01 & 0.03 & 0.15 & 0.30 &  & 134  & 3.1 & 3.1 &  & 9.9  & 8.0   &  & 7.6  & 5.9   & 6.8   \\
			&     &      & 0.10 & 0.30 & 0.80 & 0.85 &  & 464  & 3.9 & 2.0 &  & 80.0 & 0.0   &  & 82.5 & 0.0   & 73.3  \\
			&     &      & 0.10 & 0.30 & 0.15 & 0.20 &  & 101  & 3.9 & 2.6 &  & 80.0 & 2.5   &  & 83.4 & 1.2   & 73.7  \\
			&     &      & 0.10 & 0.30 & 0.15 & 0.30 &  & 134  & 3.9 & 3.9 &  & 80.0 & 31.0  &  & 82.5 & 28.9  & 74.5  \\
			&     &      &      &      &      &      &  &      &     &     &  &      &       &  &      &       &       \\
			3.5 & 1   & 3785 & 0.01 & 0.03 & 0.80 & 0.85 &  & 3142 & 3.1 & 3.3 &  & 0.0  & 0.1   &  & 0.0  & 0.1   & 0.0   \\
			&     &      & 0.01 & 0.03 & 0.15 & 0.20 &  & 681  & 3.1 & 4.1 &  & 0.0  & 52.0  &  & 0.0  & 53.2  & 39.6  \\
			&     &      & 0.01 & 0.03 & 0.15 & 0.30 &  & 908  & 3.1 & 6.1 &  & 0.0  & 100.0 &  & 0.0  & 100.0 & 100.0 \\
			&     &      & 0.10 & 0.30 & 0.80 & 0.85 &  & 3142 & 3.9 & 4.1 &  & 80.0 & 97.4  &  & 81.9 & 98.3  & 96.9  \\
			&     &      & 0.10 & 0.30 & 0.15 & 0.20 &  & 681  & 3.9 & 5.1 &  & 80.0 & 99.1  &  & 82.8 & 99.7  & 99.5  \\
			&     &      & 0.10 & 0.30 & 0.15 & 0.30 &  & 908  & 3.9 & 7.7 &  & 80.0 & 100.0 &  & 82.8 & 100.0 & 100.0 \\
			&     &      &      &      &      &      &  &      &     &     &  &      &       &  &      &       &       \\
			3.5 & 1.5 & 3785 & 0.01 & 0.03 & 0.80 & 0.85 &  & 3142 & 3.1 & 2.2 &  & 0.0  & 0.0   &  & 0.0  & 0.0   & 0.0   \\
			&     &      & 0.01 & 0.03 & 0.15 & 0.20 &  & 681  & 3.1 & 2.7 &  & 0.0  & 0.0   &  & 0.0  & 0.0   & 0.0   \\
			&     &      & 0.01 & 0.03 & 0.15 & 0.30 &  & 908  & 3.1 & 4.1 &  & 0.0  & 56.1  &  & 0.0  & 57.8  & 44.7  \\
			&     &      & 0.10 & 0.30 & 0.80 & 0.85 &  & 3142 & 3.9 & 2.7 &  & 80.0 & 0.0   &  & 83.7 & 0.0   & 73.7  \\
			&     &      & 0.10 & 0.30 & 0.15 & 0.20 &  & 681  & 3.9 & 3.4 &  & 80.0 & 5.0   &  & 83.8 & 3.1   & 75.2  \\
			&     &      & 0.10 & 0.30 & 0.15 & 0.30 &  & 908  & 3.9 & 5.1 &  & 80.0 & 99.6  &  & 82.8 & 100.0 & 99.6  \\
			&     &      &      &      &      &      &  &      &     &     &  &      &       &  &      &       &       \\
			3.5 & 2   & 3785 & 0.01 & 0.03 & 0.80 & 0.85 &  & 3142 & 3.1 & 1.6 &  & 0.0  & 0.0   &  & 0.0  & 0.0   & 0.0   \\
			&     &      & 0.01 & 0.03 & 0.15 & 0.20 &  & 681  & 3.1 & 2.0 &  & 0.0  & 0.0   &  & 0.0  & 0.0   & 0.0   \\
			&     &      & 0.01 & 0.03 & 0.15 & 0.30 &  & 908  & 3.1 & 3.1 &  & 0.0  & 0.3   &  & 0.0  & 0.0   & 0.0   \\
			&     &      & 0.10 & 0.30 & 0.80 & 0.85 &  & 3142 & 3.9 & 2.0 &  & 80.0 & 0.0   &  & 83.8 & 0.0   & 74.1  \\
			&     &      & 0.10 & 0.30 & 0.15 & 0.20 &  & 681  & 3.9 & 2.6 &  & 80.0 & 0.0   &  & 83.5 & 0.0   & 73.9  \\
			&     &      & 0.10 & 0.30 & 0.15 & 0.30 &  & 908  & 3.9 & 3.9 &  & 80.0 & 31.8  &  & 84.4 & 30.1  & 76.5   \\\hline 
	\end{tabular}}
\end{table}

\section{Application} 
\subsection{Firearms in the home and suicide risk}
In this application, suicide cases (ICD-9: E950-E959) were drawn from the 1993 National Mortality Followback Survey (NMFS) which contains a national representative sample of death certificates for US residents 15 years of age or older who died in 1993 (excluding decedents from South Dakota). Living referents were drawn from the 1994 National Health Interview Survey (NHIS) which contains a nationally representative sample of noninstitutionalized civilians 17 years of age or older in the US. 

To define the narrow cases, the information about place of suicide was obtained from the item ``Where did the (homicide/suicide/fatal accident or injury) happen?'' in the 1993 NMFS. We defined the suicide case to be a narrow case (suicide at home) if the category is ``home or private area around the home'', which unfortunately does not inform us whether it is the decedent's home or someone else's. Although ideally we would want to define suicide cases that occurred at the decedent's home as the narrow case, the current narrow case definition is still a valid definition.

\subsection{Supportive analyses}
In the matched study with referent group 1, there is a lack of balance in the missing lived alone category because no control subjects are in this category. In a supportive analysis, we conduct a matched study after removing the 2.4\% ($n=33$) cases with missing information in lived alone. Following the same matching procedure as in Section 1.2,  most of the standardized mean differences (SMDs) after matching are between -0.1 and 0.1, with a few exceptions including education=missing (SMD=0.16),  veteran=no (SMD =-0.11), and population $\geq250000$ (SMD=-0.11). The results are similar to the results reported in the main article. Specifically, comparing with and without firearms at home, the odds ratio for suicide is 3.66, and the odds ratio of committing suicide at home is 4.51. The results from the sensitivity analysis are in Figure \ref{fig: sens no missing lived alone}.

\begin{figure}[h]
	\centering
	\includegraphics[scale=0.8]{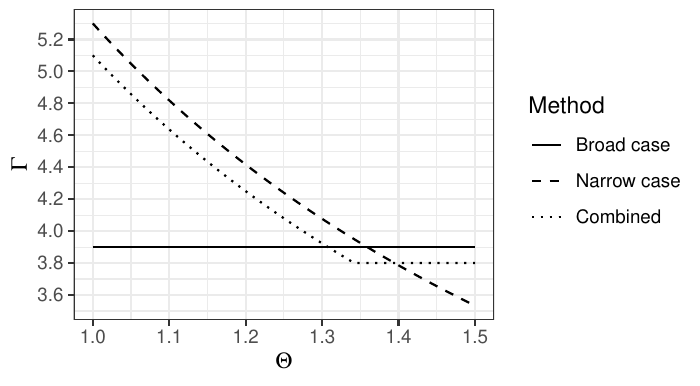}
	\caption{ The largest $ \Gamma $ against different values of the sensitivity parameter $ \Theta $, under which we are still able to conclude that there is a significant treatment effect of having firearms at home on suicide. \label{fig: sens no missing lived alone}}
\end{figure}

In another supportive analysis, we perform matching with an added caliper on the propensity score; see  \cite{rosenbaum1985constructing}. After matching, most of the standardized mean differences (SMDs) after matching are between -0.1 and 0.1, with a few exceptions including lived alone =missing (SMD=0.22),  education = missing (SMD =0.14), veteran = missing (SMD = 0.15)
and population $\geq250000$ (SMD=-0.12). Still, the results are similar to the results reported in the main article. Specifically, comparing with and without firearms at home, the odds ratio for suicide is 3.52, and the odds ratio of committing suicide at home is 4.38. The results from the sensitivity analysis are in Figure \ref{fig: sens caliper}.  

\begin{figure}[h]
	\centering
	\includegraphics[scale=0.8]{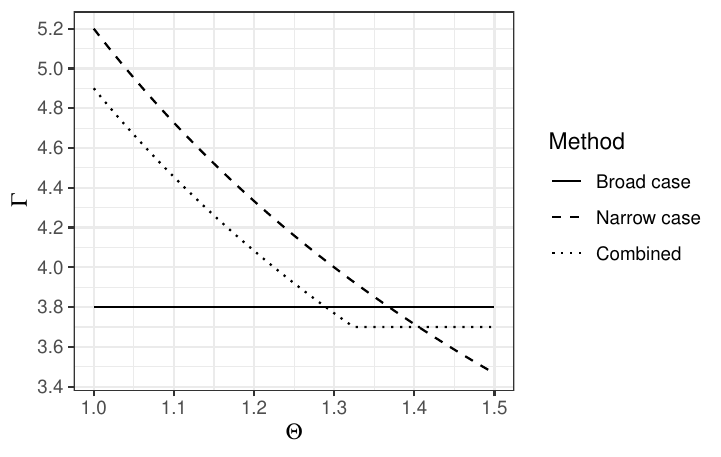}
	\caption{ The largest $ \Gamma $ against different values of the sensitivity parameter $ \Theta $, under which we are still able to conclude that there is a significant treatment effect of having firearms at home on suicide. \label{fig: sens caliper}}
\end{figure}

\section{Technical Proofs}

\subsection{Proof of Proposition \ref{proposition: Yi}} 
First note that from Bayes rule, the sensitivity model  \eqref{eq: SA}, and under $ H_0 $,   the following are true:
(i) $ \pr( Z_{ij}= 1, \kappa_n(R_{ij})= 1\mid x_{ij}, u_{ij}, \kappa_b(R_{ij}) = 1 ) = \pi_{ij}\theta_{Tij} $, \\
(ii)	$ \pr( Z_{ij}= 0, \kappa_n(R_{ij})= 1\mid x_{ij}, u_{ij}, \kappa_b(R_{ij}) = 1 ) =(1- \pi_{ij})  \theta_{Cij}$, \\
(iii) $ \pr( Z_{ij}= 1, \kappa_n(R_{ij})= 0\mid x_{ij}, u_{ij}, \kappa_b(R_{ij}) = 0 ) = \pi_{ij}$, \\ 
(iv) $ \pr( Z_{ij}= 0, \kappa_n(R_{ij})= 0\mid x_{ij}, u_{ij}, \kappa_b(R_{ij}) = 0 ) = 1- \pi_{ij}$.  \\
We can also easily calculate that $ \pr (Z_{ij}=1 \mid i\in \mathcal{N},\mathcal{F}) = \pi_{ij} $, for $ j= 2,\dots, J$, and 
\begin{align*}
	&\pr (Z_{i1}=1 \mid i\in \mathcal{N},\mathcal{F}) \\
	&= \pr (Z_{i1}=1 \mid \kappa_n(R_{i1}) = \kappa_b(R_{i1})=  1, x_{i1}, u_{i1})   \\
	&=  \frac{ \pr (Z_{i1}= 1, \kappa_n(R_{i1} )= 1\mid \kappa_b({R_{i1}} )= 1, x_{i1}, u_{i1}) }{\pr (\kappa_n(R_{i1} )= 1\mid \kappa_b({R_{i1}} )= 1, x_{i1}, u_{i1}) } \\
	&= \frac{\pi_{i1} \theta_{Ti1}}{ \pi_{i1} \theta_{Ti1}  + (1-\pi_{i1} )\theta_{Ci1} }. 
\end{align*}

If $ m_i=0 $, $ \pr \left( Y_i =1 \mid i\in \mathcal{N},  \mathcal{F}, \mathcal{Z} \right)= 0 $; if $ m_i=J $, $ \pr \left( Y_i =1 \mid i\in \mathcal{N}, \mathcal{F}, \mathcal{Z} \right)= 1 $. Under these two circumstances, Proposition   \ref{proposition: Yi} obviously holds.  For $ 1\leq m_i\leq J-1 $,   let $ \mathscr{T} (J, m_i) $ be the set containing the $ \binom{J}{m_i} $ possible values of $ \bm{Z}_i $. Then 
\begin{align*}
	&\pr \left( Y_{i} =1 \mid i\in \mathcal{N},  \mathcal{F}, \mathcal{Z} \right)  \\
	&= \pr \left( Z_{i1} =1 \mid i\in \mathcal{N},  \mathcal{F}, \mathcal{Z} \right)  \\
	&=\frac{ \pr \left( Z_{i1} =1, \bm{Z}_i\in \mathcal{Z} \mid  i\in \mathcal{N},  \mathcal{F} \right) }{  \pr \left( Z_{i1} =1, \bm{Z}_i\in \mathcal{Z} \mid  i\in \mathcal{N},  \mathcal{F} \right)+  \pr \left( Z_{i1} =0, \bm{Z}_i\in \mathcal{Z} \mid  i\in \mathcal{N},  \mathcal{F} \right)} \\
	& =\frac{ \sum_{\bm{z} \in  \mathscr{T} (J-1, m_i-1) } 	  \pr \left( Z_{i1} =1, \bm{Z}_{i, -1} = \bm{z} \mid  i\in \mathcal{N},  \mathcal{F}\right)  }{ \sum_{\bm{z} \in  \mathscr{T} (J-1, m_i-1) } 	  \pr \left( Z_{i1} =1, \bm{Z}_{i, -1} = \bm{z} \mid  i\in \mathcal{N},  \mathcal{F}\right)  + \sum_{\bm{z} \in  \mathscr{T} (J-1, m_i) } 	  \pr \left( Z_{i1} =0, \bm{Z}_{i, -1} = \bm{z} \mid i\in \mathcal{N},  \mathcal{F}\right)  }\\
	& =\frac{ \pi_{i1} \theta_{Ti1}   \sum_{\bm{z} \in  \mathscr{T} (J-1, m_i-1) } \prod_{j=2}^{J} 
		\pi_{ij}^{z_{ij}}	 (1-	\pi_{ij})^{1-z_{ij}}	  }{ \pi_{i1} \theta_{Ti1}   \sum_{\bm{z} \in  \mathscr{T} (J-1, m_i-1) } \prod_{j=2}^{J} 
		\pi_{ij}^{z_{ij}}	 (1-	\pi_{ij})^{1-z_{ij}}	   + (1-\pi_{i1}) \theta_{Ci1}   \sum_{\bm{z} \in  \mathscr{T} (J-1, m_i) } \prod_{j=2}^{J} 
		\pi_{ij}^{z_{ij}}	 (1-	\pi_{ij})^{1-z_{ij}}	    }\\
	& =\frac{  \theta_{Ti1}  \exp(\log (\Gamma)u_{i1})  \sum_{\bm{z} \in  \mathscr{T} (J-1, m_i-1) } \exp (\log (\Gamma) \sum_{j=2}^J z_{ij} u_{ij} )	  }{  \theta_{Ti1}  \exp(\log (\Gamma)u_{i1})  \sum_{\bm{z} \in  \mathscr{T} (J-1, m_i-1) } \exp (\log (\Gamma) \sum_{j=2}^J z_{ij} u_{ij} )    +  \theta_{Ci1}  \sum_{\bm{z} \in  \mathscr{T} (J-1, m_i) } \exp (\log (\Gamma) \sum_{j=2}^J z_{ij} u_{ij} )   }. 
\end{align*}

Write $ \omega_{ij} = \exp \left( \log (\Gamma) u_{ij}\right) >0$ and $ \bm \omega_i = (\omega_{i2}, \dots, \omega_{iJ})^T $; note that  $  \bm \omega_i  $ excludes $ \omega_{i1}  $. Then for each integer $ b $, $ 0\leq b\leq J-1 $, 
\begin{align*}
	\sum_{\bm{z} \in  \mathscr{T} (J-1, b) } \exp \left(\log (\Gamma) \sum_{j=2}^J z_{ij} u_{ij} \right) =  	 \sum_{\bm{z} \in  \mathscr{T} (J-1, b) }  \prod_{j=2}^J \omega_{ij}^{z_{ij}} : = S_b (\bm\omega_i),  
\end{align*} 
where $ S_b (\bm\omega_i)  $ is the $ b $th elementary symmetric function of $ \bm\omega_i $. Using $ S_b (\bm\omega_i) $, we can write 
\begin{align}
	\pr \left( Y_{i} =1 \mid i\in \mathcal{N},  \mathcal{F}, \mathcal{Z} \right)  &= \frac{  \theta_{Ti1}  \exp(\log (\Gamma)u_{i1})  S_{m_i- 1} (\bm\omega_i)	  }{  \theta_{Ti1}  \exp(\log (\Gamma)u_{i1}) S_{m_i- 1} (\bm\omega_i)	   +  \theta_{Ci1}  S_{m_i} (\bm\omega_i)	  }\nonumber\\
	&= \frac{  \frac{\theta_{Ti1}  }{\theta_{Ci1} }\exp(\log (\Gamma)u_{i1})  }{  \frac{\theta_{Ti1}  }{\theta_{Ci1} } \exp(\log (\Gamma)u_{i1}) 	   +   \frac{S_{m_i} (\bm\omega_i)	 }{  S_{m_i- 1} (\bm\omega_i)	 } } \label{eq: supp pYi}. 
\end{align}
Note that \eqref{eq: supp pYi}  is clearly  monotone increasing in $ \frac{\theta_{Ti1}  }{\theta_{Ci1} }\exp(\log (\Gamma)u_{i1})  $. Also, from a property of elementary symmetric functions \cite[Section 2.15.3, Theorem 1, page 102]{mitrinovic1970analytic}, $  \frac{S_{m_i} (\bm\omega_i)	 }{  S_{m_i- 1} (\bm\omega_i)	 }  $ is increasing in each $ w_{ij} $, and thus also in each $ u_{ij} $,  $ 2\leq j\leq J $. So $ \pr \left( Y_{i} =1 \mid i\in \mathcal{N},  \mathcal{F}, \mathcal{Z} \right)   $  is maximized  at $ \bm{u}_i= (1,0 \dots, 0) $ and  $ \theta_{Ti1}/\theta_{Ci1} = \Theta $;  minimized at $ \bm{u}_i= (0,1 \dots, 1) $ and $ \theta_{Ti1}/\theta_{Ci1} = 1 $. When $ \bm{u}_i= (1,0 \dots, 0) $, $ S_{b} (\bm \omega_i ) = \binom{J-1}{b}$ and   $ S_{m_i} (\bm \omega_i ) / S_{m_i-1} (\bm \omega_i ) =  \binom{J-1}{m_i}/ \binom{J-1}{m_i-1} = (J- m_i)/m_i$; when $ \bm{u}_i= (0,1 \dots, 1) $, $ S_{b} (\bm \omega_i ) = \binom{J-1}{b}\Gamma^b$ and   $ S_{m_i} (\bm \omega_i ) / S_{m_i-1} (\bm \omega_i ) = \Gamma \binom{J-1}{m_i}/ \binom{J-1}{m_i-1} =\Gamma (J- m_i)/m_i$.  Hence, 
\[
\frac{m_i}{m_i+ (J-m_i)\Gamma }\leq \pr \left( Y_{i} =1 \mid i\in \mathcal{N}, \mathcal{F}, \mathcal{Z} \right)\leq \frac{m_i \Theta\Gamma}{ m_i \Theta\Gamma +  J-m_i
}.
\]

\subsection{Additional Details on \eqref{eq: power b}-\eqref{eq: power n} and Derivation of $ \tilde{\Gamma}_b, \tilde{\Gamma}_n $}
\label{subsec: additional details}
(i) For the broad case definition, after some algebra, we have 
\begin{align*}
	& \pr(m_i=0\mid  \mathcal{S} ) =\frac{(1-\pi)^J b_C(1-b_C)^{J-1} }{\{b_T \pi + b_C (1-\pi)\}\{(1-b_T)\pi + (1-b_C)(1-\pi)  \}^{J-1}},
\end{align*}
for $ t=1,\dots, J-1 $,
\begin{align*}
	& \pr(m_i=t\mid \mathcal{S} ) = \frac{\pi^t (1-\pi)^{J-t} \left\{b_T\binom{J-1}{t-1}(1-b_T)^{t-1}(1-b_C)^{J-t} + b_C \binom{J-1}{t} (1-b_T)^t (1-b_C)^{J-1-t}    \right\}}{\{b_T \pi + b_C (1-\pi)\}\{(1-b_T)\pi + (1-b_C)(1-\pi)  \}^{J-1}},
\end{align*}
and
\begin{align*}
	& \pr(m_i=J\mid \mathcal{S} ) =  \frac{\pi^J b_T (1-b_T)^{J-1}  }{\{b_T \pi + b_C (1-\pi)\}\{(1-b_T)\pi + (1-b_C)(1-\pi)  \}^{J-1}}.
\end{align*}
We thus have explicit  expressions  of $ \E (\bar{\bar p}_{bi} \mid \mathcal{S}) $ and $ \E\{\bar{\bar p}_{bi} (1- \bar{\bar p}_{bi}) \mid \mathcal{S}\} $ using the probability mass function of $ m_i $ conditional on $ \mathcal{S} $.  

It remains to derive the analytical form of  the design sensitivity $ \tilde{\Gamma}_b $, that solves 
\begin{align}
	\frac{b_T\pi }{b_T\pi + b_C (1-\pi )} = \E\left(\frac{m_i \Gamma}{m_i \Gamma+ J-m_i}\mid \mathcal{S} \right) .\nonumber
\end{align} 
From some algebra, we can calculate that 
\begin{align*}
	&\E\left(\frac{m_i \tilde \Gamma_b}{m_i \tilde\Gamma_b+ J-m_i}\mid \mathcal{S} \right)  \\
	&\qquad = \sum_{t=1}^{J-1}  \frac{t \tilde\Gamma_b}{t \tilde \Gamma_b + J-t} \frac{\pi^t (1-\pi )^{J-t} \left\{ b_T  \binom{J-1}{t-1} 
		(1- b_T)^{t-1} (1-b_C)^{J-t} + b_C \binom{J-1}{t} (1-b_T)^t (1-b_C)^{J-1-t}   	\right\} }{\{b_T \pi + b_C (1-\pi)\}\{(1-b_T)\pi + (1-b_C)(1-\pi)  \}^{J-1} }  \\
	&\qquad \qquad +  1\cdot  \frac{\pi^J b_T(1-b_T)^{J-1} }{ \{b_T \pi + b_C (1-\pi)\}\{(1-b_T)\pi + (1-b_C)(1-\pi)  \}^{J-1} } . 
\end{align*}
Equating this to $ 	\frac{b_T\pi }{b_T\pi + b_C (1-\pi )} $, we can derive that 
\begin{align*}
	&\sum_{t=1}^{J-1}  \frac{t \tilde\Gamma_b}{t \tilde \Gamma_b + J-t}  \pi^t (1-\pi )^{J-t} \left\{ b_T  \binom{J-1}{t-1} 
	(1- b_T)^{t-1} (1-b_C)^{J-t} + b_C \binom{J-1}{t} (1-b_T)^t (1-b_C)^{J-1-t}   	\right\}  \\
	&\qquad = b_T \pi  \{(1-b_T)\pi + (1-b_C)(1-\pi)\}^{J-1}  - \pi^J b_T (1-b_T)^{J-1} \\
	&\qquad = b_T\pi \sum_{t=0}^{J-1} \binom{J-1}{t} (1-b_T)^t\pi^t (1-b_C)^{J-1-t}    (1-\pi )^{J-1-t} - \pi^J b_T (1-b_T)^{J-1} \\
	&\qquad = b_T\pi \sum_{s=1}^{J} \binom{J-1}{s-1} (1-b_T)^{s-1}\pi^{s-1} (1-b_C)^{J-s}    (1-\pi )^{J-s} - \pi^J b_T (1-b_T)^{J-1} \\
	&\qquad = \sum_{s=1}^{J-1} b_T \binom{J-1}{s-1} (1-b_T)^{s-1}\pi^{s} (1-b_C)^{J-s}    (1-\pi )^{J-s}  + b_T\pi(1-b_T)^{J-1}\pi^{J-1}- \pi^J b_T (1-b_T)^{J-1} \\
	&\qquad =  \sum_{s=1}^{J-1} b_T \binom{J-1}{s-1} (1-b_T)^{s-1}\pi^{s} (1-b_C)^{J-s}    (1-\pi )^{J-s}  . 
\end{align*}
Arranging the terms and collecting all the terms inside the summation, we have that 
\begin{align*}
	&\sum_{t=1}^{J-1}  \bigg\{\frac{J-t}{t \tilde \Gamma_b + J-t}  \pi^t (1-\pi )^{J-t} b_T  \binom{J-1}{t-1} 
	(1- b_T)^{t-1} (1-b_C)^{J-t}  \\
	&\qquad \qquad - \frac{t\tilde{\Gamma}_b}{t\tilde{\Gamma}_b + J-t} \pi^t(1-\pi)^{J-t} b_C \binom{J-1}{t} (1-b_T)^t (1-b_C)^{J-1-t}    \bigg\}= 0,
\end{align*}
which gives 
\begin{align*}
	& \bigg\{ \sum_{t=1}^{J-1} \frac{t}{t \tilde \Gamma_b + J-t}  \pi^t (1-\pi )^{J-t}   \binom{J-1}{t} 
	(1- b_T)^{t-1} (1-b_C)^{J-t-1}  \bigg\}   (1-b_C)b_T\\
	& = \bigg\{   \sum_{t=1}^{J-1} \frac{t}{t\tilde{\Gamma}_b + J-t} \pi^t(1-\pi)^{J-t} \binom{J-1}{t} (1-b_T)^{t-1} (1-b_C)^{J-1-t}    \bigg\} \tilde{\Gamma}_b b_C(1-b_T) . 
\end{align*}
Then we immediately have the expression of $ \tilde\Gamma_b $ as
\begin{align*}
	\tilde\Gamma_b= \frac{b_T/(1-b_T)}{b_C/(1-b_C)}. 
\end{align*}

(ii) Similarly for the narrow case definition, we have  
\begin{align*}
	& \pr(m_i=0\mid \mathcal{S}, i\in \mathcal{N}) =\frac{(1-\pi)^J b_C\eta_C(1-b_C)^{J-1} }{\{b_T\eta_T \pi + b_C\eta_C (1-\pi)\}\{(1-b_T)\pi + (1-b_C)(1-\pi)  \}^{J-1}},
\end{align*}
for $ t=1,\dots, J-1 $,
\begin{align*}
	& \pr(m_i=t\mid \mathcal{S}, i\in \mathcal{N}) \\
	=& \frac{\pi^t (1-\pi)^{J-t} \left\{b_T\eta_T\binom{J-1}{t-1}(1-b_T)^{t-1}(1-b_C)^{J-t} + b_C\eta_C \binom{J-1}{t} (1-b_T)^t (1-b_C)^{J-1-t}    \right\}}{\{b_T \eta_T \pi + b_C\eta_C (1-\pi)\}\{(1-b_T)\pi + (1-b_C)(1-\pi)  \}^{J-1}},
\end{align*}
and 
\begin{align*}
	& \pr(m_i=J\mid  \mathcal{S}, i\in \mathcal{N}) = \frac{\pi^J  b_T\eta_T (1-b_T)^{J-1} }{\{b_T \eta_T \pi + b_C\eta_C (1-\pi)\}\{(1-b_T)\pi + (1-b_C)(1-\pi)  \}^{J-1}}.
\end{align*}
We thus have explicit  expressions  of $ \E (\bar{\bar p}_{ni} \mid \mathcal{S}, i\in \mathcal{N}) $ and $ \E\{\bar{\bar p}_{bi} (1- \bar{\bar p}_{bi}) \mid \mathcal{S}, \mathcal{N}\} $ using the probability mass function of $ m_i $ conditional on $ \mathcal{S} , i \in \mathcal{N}$.

Furthermore, the design sensitivity $ \tilde{\Gamma}_n $ solves 
\begin{align}
	\frac{b_T\eta_T \pi }{b_T\eta_T\pi + b_C\eta_C (1-\pi )} = \E\left(\frac{m_i\Theta \Gamma}{m_i\Theta \Gamma+ J-m_i}\mid \mathcal{S}, i\in \mathcal{N} \right) .\nonumber
\end{align} 
The expression of $ \tilde\Gamma_n $ follows similar  steps as the derivation of  $ \tilde\Gamma_b $. For completeness, we lay out the steps as follows. Note that 
\begin{align*}
	&\E\left(\frac{m_i \Theta \tilde \Gamma_n}{m_i \Theta\tilde \Gamma_n+ J-m_i}\mid \mathcal{S}, i\in \mathcal{N} \right)  \\
	& = \sum_{t=1}^{J-1}  \frac{t \Theta \tilde \Gamma_n}{t \Theta \tilde \Gamma_n + J-t} \frac{\pi^t (1-\pi )^{J-t} \left\{ b_T\eta_T  \binom{J-1}{t-1} 
		(1- b_T)^{t-1} (1-b_C)^{J-t} + b_C \eta_C\binom{J-1}{t} (1-b_T)^t (1-b_C)^{J-1-t}   	\right\} }{\{b_T \eta_T\pi + b_C \eta_C (1-\pi)\}\{(1-b_T)\pi + (1-b_C)(1-\pi)  \}^{J-1} }  \\
	&\qquad \qquad +  1\cdot  \frac{\pi^J b_T\eta_T (1-b_T)^{J-1} }{ \{b_T\eta_T \pi + b_C \eta_C(1-\pi)\}\{(1-b_T)\pi + (1-b_C)(1-\pi)  \}^{J-1} } . 
\end{align*}
Equating this to $ 	\frac{b_T\eta_T \pi }{b_T\eta_T\pi + b_C\eta_C (1-\pi )}  $, we can derive that  
\begin{align*}
	&\sum_{t=1}^{J-1}  \frac{t \Theta \tilde\Gamma_n}{t \Theta \tilde \Gamma_n + J-t}  \pi^t (1-\pi )^{J-t} \left\{ b_T \eta_T \binom{J-1}{t-1} 
	(1- b_T)^{t-1} (1-b_C)^{J-t} + b_C \eta_C \binom{J-1}{t} (1-b_T)^t (1-b_C)^{J-1-t}   	\right\}  \\
	&\qquad = b_T \eta_T \pi  \{(1-b_T)\pi + (1-b_C)(1-\pi)\}^{J-1}  - \pi^J b_T\eta_T (1-b_T)^{J-1} \\
	&\qquad = b_T\eta_T \pi \sum_{t=0}^{J-1} \binom{J-1}{t} (1-b_T)^t\pi^t (1-b_C)^{J-1-t}    (1-\pi )^{J-1-t} - \pi^J b_T\eta_T (1-b_T)^{J-1} \\
	&\qquad = b_T\eta_T\pi \sum_{s=1}^{J} \binom{J-1}{s-1} (1-b_T)^{s-1}\pi^{s-1} (1-b_C)^{J-s}    (1-\pi )^{J-s} - \pi^J b_T\eta_T (1-b_T)^{J-1} \\
	&\qquad = b_T\eta_T \sum_{s=1}^{J-1}  \binom{J-1}{s-1} (1-b_T)^{s-1}\pi^{s} (1-b_C)^{J-s}    (1-\pi )^{J-s}  . 
\end{align*}
Arranging the terms and collecting all the terms inside the summation, we have that 
\begin{align*}
	&\bigg\{ \sum_{t=1}^{J-1}  \frac{J-t}{t \Theta\tilde \Gamma_n + J-t}  \pi^t (1-\pi )^{J-t}  \binom{J-1}{t-1} 
	(1- b_T)^{t-1} (1-b_C)^{J-1- t}  \bigg\} b_T  \eta_T (1-b_C)\\
	&=\bigg\{ \sum_{t=1}^{J-1}  \frac{J-t }{t\Theta\tilde{\Gamma}_n + J-t} \pi^t(1-\pi)^{J-t} \binom{J-1}{t-1} (1-b_T)^{t-1} (1-b_C)^{J-1-t}    \bigg\}\Theta \tilde \Gamma_n  b_C\eta_C (1-b_T). 
\end{align*}
Then we immediately have the expression of $ \tilde\Gamma_n $ as
\begin{align*}
	\tilde\Gamma_n= \frac{b_T/(1-b_T)}{b_C/(1-b_C)}\times \frac{\eta_T/\eta_C}{\Theta}. 
\end{align*}

\end{document}